\documentclass[12pt,a4paper]{article}
\usepackage{graphicx}
\usepackage{amsmath}
\usepackage{amssymb}
\usepackage{listings} 

\pdfoutput=1

\newcommand{\asb}{\bar{\alpha}_s}

\newcommand{\stringa}{\ttfamily\lstinline}
\def\cod#1{{\stringa!#1!}}

\title{\bf Multi-Regge kinematics and azimuthal angle observables for inclusive four-jet production} 
\author{F. Caporale$^1$, F.~G. Celiberto$^{1,2}$, G. Chachamis$^1$, A. Sabio Vera$^{1}$\\ \\
{\small $^1$ Instituto de F{\' \i}sica Te{\' o}rica UAM/CSIC, Nicol{\'a}s Cabrera 15}\\ 
{\small \& Universidad Aut{\' o}noma de Madrid, E-28049 Madrid, Spain.}\\
{\small $^2$ Dipartimento di Fisica, Universit{\`a} della Calabria \&}\\
{\small Istituto Nazionale di Fisica Nucleare, Gruppo Collegato di Cosenza,}\\
{\small I-87036 Arcavacata di Rende, Cosenza, Italy.}
}

\begin{document} 

\maketitle 

\abstract

We evaluate differential cross sections for production of four jets in multi-Regge kinematics at a hadron collider. The main focus lies on azimuthal angle dependences. As in previous studies, the ratios of correlation functions of products of cosines 
of azimuthal angle differences among the tagged jets offer us the cleanest quantities to compare with experimental data. The calculations are based on the jet production from a single BFKL ladder with a convolution of three BFKL Green functions where we always have two forward/backward jets tagged in the final state. We also demand the tagging of two further jets in more central regions of the detectors with a relative separation in rapidity from each other, plus the inclusive production of an arbitrary number of mini-jets. 
We show that dependences on the transverse momenta and rapidity of the two central jets can be a distinct signal of the onset of BFKL dynamics.  

\section{Introduction}

The study of the asymptotic behavior of scattering amplitudes in the limit of high center-of-mass energy is an active area of research for particle phenomenology. The Large Hadron Collider (LHC) 
is producing an abundance of data allowing for the study of very exclusive observables with stringent cuts in the final state. One of the key points for understanding
multi-jet production at high energies is the multi-Regge kinematics (MRK). MRK is the kinematics
that, by presupposing a strong ordering in rapidity for the final state jets,
allows for large logarithms in the colliding energy to be present in all orders of the perturbative
expansion. This fact alone calls for a resummation framework of the large logarithms in energy.

In the high energy (Regge) limit, the common basis for the perturbative description of a hard process in QCD is the Balitsky-Fadin-Kuraev-Lipatov (BFKL) approach, at leading (LL) \cite{Lipatov:1985uk,Balitsky:1978ic,Kuraev:1977fs,Kuraev:1976ge,Lipatov:1976zz,Fadin:1975cb} 
and next-to-leading (NLL)~\cite{Fadin:1998py,Ciafaloni:1998gs} accuracy.
This approach offers a resummation of those enhanced terms in MRK in regions of phase space where a fixed order calculation might not be enough. This formalism has been successfully applied to lepton-hadron Deep Inelastic Scattering at HERA (see, {\it e.g.}~ \cite{Hentschinski:2012kr,Hentschinski:2013id}) to describe quite inclusive processes which are not that suitable though if one is
interested in discriminating between BFKL dynamics and other resummation programs. At the LHC, however, it is possible to investigate processes with much more exclusive final states which could, in principle, be only described by the BFKL framework. This would allow us to precisely determine the 
applicability window of the framework. 

With this idea in mind there has been a lot of recent activity in the study of the so-called Mueller-Navelet jets~\cite{Mueller:1986ey}, {\it i.e.} the inclusive hadro-production of two forward jets with large and similar transverse momenta and a big relative separation in rapidity $Y$, proportional to $\sqrt{s}$, and with associated inclusive mini-jet radiation. Interesting observables associated to this process are the azimuthal angle ($\theta$) correlations $\langle \cos{(n \, \theta)} \rangle$ of the two tagged jets, and it has been shown~\cite{DelDuca:1993mn,Stirling:1994he,Orr:1997im,Kwiecinski:2001nh} that the further gluon radiation manifests as a fast decrease of these functions with $Y$. However, these observables are strongly affected by collinear effects~\cite{Vera:2006un,Vera:2007kn} due to their dependence on the $n=0$ Fourier component in $\theta$ of the BFKL kernel, which is strongly dependent on collinear radiation. In order to remove this problem, new observables were proposed~\cite{Vera:2006un,Vera:2007kn} which are independent from the $n=0$ contribution: the ratios 
${\cal R}^{M}_{N} = \langle \cos{(M \, \theta)} \rangle / \langle \cos{(N \, \theta)} \rangle$.  They have been calculated at NLL~\cite{Ducloue:2013bva,Caporale:2014gpa,
Celiberto:2015dgl,Ciesielski:2014dfa,Angioni:2011wj}  
and show a very good agreement with experimental data at the LHC. 

Nevertheless, Mueller-Navelet configurations are still too inclusive to study MRK with precision. 
A step toward other observables capable to pin down the MRK dynamics in much more detail
has been taken in~\cite{Caporale:2015vya},
where a new study was proposed that demands the tagging of a third, central in rapidity, jet 
within the usual Mueller-Navelet configuration.
It is important to remain within the general Mueller-Navelet setup because having two forward/backward jets allows the use of collinear factorization which is in a better theoretical control than $k_t$-factorization. 
Since a unique footprint of BFKL physics is its azimuthal angle dependence, the main new observables studied in~\cite{Caporale:2015vya} are the ratios  
$${\cal R}^{M N}_{P Q} =\frac{ \langle \cos{(M \, \phi_1)}  \cos{(N \, \phi_2)}  \rangle}{ \langle \cos{(P \, \phi_1)}  \cos{(Q \, \phi_2)} \rangle} \, ,$$ 
with $\phi_1$ and $\phi_2$ being the azimuthal angle difference respectively between the first and the second (central) jet and between this one and the third jet. These observables depend strongly on the $p_t$ and less strongly on the rapidity of the central jet, and this information can be used to probe characteristic properties of the BFKL ladder in a very precise way. 

The present work, original results of which are presented in the next Section,  is a natural continuation of~\cite{Caporale:2015vya}, by allowing the production of a second central jet, thus making it possible to define more differential distributions in the transverse momenta, 
azimuthal angles and rapidities of the two central jets, for fixed values 
of the four momenta of the forward jets.

\section{Inclusive four-jet production}

We now present the analysis of events with  two forward/backward jets together with two more central jets, all of them well separated in rapidity from each other,
making use of the BFKL formalism to describe the associated inclusive multi-jet emission.  
The two tagged forward/backward jets $A$ and $B$ 
have transverse momentum $\vec{k}_{A,B}$, azimuthal angle 
$\vartheta_{A,B}$ and rapidity $Y_{A,B}$, while the pair of tagged more central jets are characterized,  respectively, by $\vec{k}_{1,2}$, $\vartheta_{1,2}$ and $y_{1,2}$. The differential cross section 
on these latter variables can be written in the form
\begin{align}\label{d6sigma}
 & \hspace{-0.5cm}
 \frac{d^6\sigma^{\rm 4-jet} \left(\vec{k_A},\vec{k_B},Y_A-Y_B\right)}
      {d^2\vec{k_1} dy_1 d^2\vec{k_2} dy_2}
 \\ & \nonumber \hspace{-0.15cm} 
 = 
 \frac{\asb^2}{\pi^2 k_1^2 k_2^2}
 \int d^2\vec{p_A} \int d^2\vec{p_B} \int d^2\vec{p_1} \int d^2\vec{p_2}
 \\ & \nonumber \hspace{0.40cm}
 \delta^{(2)}\left(\vec{p_A}+\vec{k_1}-\vec{p_1}\right)
 \delta^{(2)}\left(\vec{p_B}-\vec{k_2}-\vec{p_2}\right)
 \\ & \nonumber \hspace{0.40cm}
 \varphi\left(\vec{k_A},\vec{p_A},Y_A-y_1\right)
 \varphi\left(\vec{p_1},\vec{p_2},y_1-y_2\right)
 \varphi\left(\vec{p_B},\vec{k_B},y_2-Y_B\right).
\end{align}
Here we have introduced the rapidity ordering characteristic of MRK: $Y_A > y_1 > y_2 > Y_B$; and $k_1^2$, $k_2^2$ lie above the experimental resolution scale. 
$\varphi$ are BFKL gluon Green functions normalized 
to $ \varphi \left(\vec{p},\vec{q},0\right) = \delta^{(2)} \left(\vec{p} - \vec{q}\right)/(2 \pi)$ and $\bar{\alpha}_s = \alpha_s N_c/\pi$.

Following the course taken in Ref.~\cite{Caporale:2015vya}, our goal is to define and study 
the behavior of observables for which the BFKL approach 
will show distinct features with respect to other formalisms and, if possible, are also quite insensitive to higher-order corrections. We start with the study of a quantity similar to the usual Mueller-Navelet case such that we integrate over the azimuthal angles 
of the two central jets and over the difference in azimuthal angle 
between the two forward jets, $\Delta\theta = \vartheta_A - \vartheta_B - \pi$, 
to define
\begin{align} \label{Dth_th1th2_d6sigma}
 &  \hspace{-0.30cm}
 \int_0^{2\pi} d\Delta\theta \cos\left(M\Delta\theta\right)
 \int_0^{2\pi} d\vartheta_1 \int_0^{2\pi} d\vartheta_2
 \frac{d^6\sigma^{\rm 4-jet}\left(\vec{k_A},\vec{k_B},Y_A-Y_B\right)}
      {dk_1 dy_1 d\vartheta_1 dk_2 d\vartheta_2 dy_2}     
 \\ & \nonumber \hspace{0.50cm}
 =
 \frac{4 \asb^2}{k_1 k_2}
  \left(e^{iM\pi} \, 
   \tilde{\Omega}_{M}(\vec{k_A},\vec{k_B},Y_A,Y_B,\vec{k_1},\vec{k_2},
                           y_1,y_2)
   +    c.c.
  \right)
\end{align}
where 
\begin{align}\label{omega_n_tilde}
 &    
 \tilde{\Omega}_{n}(\vec{k_A},\vec{k_B},Y_A,Y_B,\vec{k_1},\vec{k_2},
                    y_1,y_2)
 \\ & \nonumber
 =
 \int_0^{+\infty} dp_A \, p_A \int_0^{+\infty} dp_B \, p_B
 \int_0^{2\pi} d\phi_A \int_0^{2\pi} d\phi_B
 \\ & \nonumber
 \frac{\left(p_A+k_1 e^{-i\phi_A}\right)^n \,
       \left(p_B-k_2 e^{i\phi_B}\right)^n}
      {
       \sqrt{\left(p_A^2+k_1^2+2 p_A k_1 \cos\phi_A\right)^n}
       \
       \sqrt{\left(p_B^2+k_2^2-2 p_B k_2 \cos\phi_B\right)^n}
      }
 \\ & \nonumber
 \varphi_n\left(|\vec{k_A}|,|\vec{p_A}|,Y_A-y_1\right)
 \varphi_n\left(|\vec{p_B}|,|\vec{k_B}|,y_2-Y_B\right)
 \\ & \nonumber
 \varphi_n\left
   (\sqrt{p_A^2+k_1^2+2 p_A k_1 \cos\phi_A}
   ,\sqrt{p_B^2+k_2^2-2 p_B k_2 \cos\phi_B}
   ,y_1-y_2\right)
\end{align}
and 
\begin{align}
 \varphi_{n} \left(|p|,|q|,Y\right) \; &= \; 
 \int_0^\infty d \nu   
 \cos{\left(\nu \ln{\frac{p^2}{q^2}}\right)}  
 \frac{e^{\bar{\alpha}_s  \chi_{|n|} \left(\nu\right) Y}}
      {\pi^2 \sqrt{p^2 q^2} }, \label{phin}
 \\
 \chi_{n} \left(\nu\right) \; &= \; 2\, \psi (1) - 
 \psi \left( \frac{1+n}{2} + i \nu\right) - 
 \psi \left(\frac{1+n}{2} - i \nu\right)
\end{align}
($\psi$ is the logarithmic derivative of Euler's gamma function).
The associated experimental observable corresponds to the mean value
of the cosine of $\Delta\theta = \vartheta_A - \vartheta_B - \pi$ 
in the recorded events:
\begin{align}\label{C0}
 & \hspace{-0.50cm}
 \left\langle\cos(M(\vartheta_A - \vartheta_B - \pi))\right\rangle
 \\ & \nonumber \hspace{0.30cm}
 =
 \frac{\int_0^{2\pi} d\Delta\theta \cos(M\Delta\theta)
       \int_0^{2\pi} d\vartheta_1 \int_0^{2\pi} d\vartheta_2
       \frac{d^6\sigma^{\rm 4-jet}}
            {dk_1 dy_1 d\vartheta_1 dk_2 d\vartheta_2 dy_2}}
      {\int_0^{2\pi} d\Delta\theta 
       \int_0^{2\pi} d\vartheta_1 \int_0^{2\pi} d\vartheta_2
       \frac{d^6\sigma^{\rm 4-jet}}
            {dk_1 dy_1 d\vartheta_1 dk_2 d\vartheta_2 dy_2}}.
 \end{align}
In order to improve the perturbative stability of our predictions  
(see~\cite{Caporale:2013uva} for a related discussion) it is convenient to remove 
the contribution from the zero conformal spin (which corresponds to the index $n=0$ in Eq.~(\ref{phin}))  by defining the ratios
\begin{equation}\label{Rmn}
 \mathcal{R}^M_N =
 \frac{\left\langle\cos(M(\vartheta_A - \vartheta_B - \pi))\right\rangle}
      {\left\langle\cos(N(\vartheta_A - \vartheta_B - \pi))\right\rangle}
\end{equation}
where we consider $M,N$ as positive integers. 

Our next step now is to propose new observables, 
different from those characteristic of the Mueller-Navelet case though still related 
to azimuthal angle projections. We thus define
\begin{align}\label{projections_1} 
 \mathcal{C}_{MNL} = & 
 \int_0^{2\pi} d\vartheta_A \int_0^{2\pi} d\vartheta_B
 \int_0^{2\pi} d\vartheta_1 \int_0^{2\pi} d\vartheta_2
 \\ & \nonumber
 \cos\left(M\left(\vartheta_A-\vartheta_1-\pi\right)\right)
 \cos\left(N\left(\vartheta_1-\vartheta_2-\pi\right)\right)
 \\ & \nonumber
 \cos\left(L\left(\vartheta_2-\vartheta_B-\pi\right)\right)
 \frac{d^6\sigma^{\rm 4-jet}\left(\vec{k_A},\vec{k_B},Y_A-Y_B\right)}
 {dk_1 dy_1 d\vartheta_1 dk_2 d\vartheta_2 dy_2}\,,
 \end{align}
where we consider $M$, $N$, $L > 0$ and integer.
After a bit of algebra we have
\begin{align}\label{projections_3}
 & \mathcal{C}_{MNL} = 
 \frac{2\pi^2 \asb^2}{k_1 k_2} \, (-1)^{M + N + L} \:
      (  \tilde{\Omega}_{M,N,L} + \tilde{\Omega}_{M,N,-L} +
          \tilde{\Omega}_{M,-N,L}  
 \\     & \hspace{0.2cm} \nonumber + \tilde{\Omega}_{M,-N,-L} +
          \tilde{\Omega}_{-M,N,L} + \tilde{\Omega}_{-M,N,-L} +
          \tilde{\Omega}_{-M,-N,L} + \tilde{\Omega}_{-M,-N,-L})
 \end{align}
with
 \begin{align}\label{omega_mnl_tilde}
 &
 \tilde{\Omega}_{m,n,l}
 = 
 \int_0^{+\infty} dp_A \, p_A \int_0^{+\infty} dp_B \, p_B
 \int_0^{2\pi} d\phi_A \int_0^{2\pi} d\phi_B
 \\ & \nonumber
 \: \frac{e^{-im\phi_A} \, e^{il\phi_B} \,
       \left(p_A e^{i\phi_A}+k_1\right)^n \,
       \left(p_B e^{-i\phi_B}-k_2\right)^n}
      {
       \sqrt{\left(p_A^2+k_1^2+2 p_A k_1 \cos\phi_A\right)^n}
       \
       \sqrt{\left(p_B^2+k_2^2-2 p_B k_2 \cos\phi_B\right)^n}
      }
 \\ & \nonumber
 \varphi_m\left(|\vec{k_A}|,|\vec{p_A}|,Y_A-y_1\right)
  \varphi_l\left(|\vec{p_B}|,|\vec{k_B}|,y_2-Y_B\right) 
 \\ & \nonumber
 \varphi_n\left
   (\sqrt{p_A^2+k_1^2+2 p_A k_1 \cos\phi_A}
   ,\sqrt{p_B^2+k_2^2-2 p_B k_2 \cos\phi_B}
   ,y_1-y_2\right).
 \end{align}
In order to drastically reduce the dependence on collinear configurations we can remove the zero conformal spin contribution by defining the following ratios:
\begin{align}\label{R^mnl_pqr}
 &
 \mathcal{R}^{MNL}_{PQR}
 \\ & \nonumber
 =
 \frac{\left\langle\cos(M(\vartheta_A - \vartheta_1 - \pi))
                   \cos(N(\vartheta_1 - \vartheta_2 - \pi))
                   \cos(L(\vartheta_2 - \vartheta_B - \pi))\right\rangle}
      {\left\langle\cos(P(\vartheta_A - \vartheta_1 - \pi))
                   \cos(Q(\vartheta_1 - \vartheta_2 - \pi))
                   \cos(R(\vartheta_2 - \vartheta_B - \pi))\right\rangle}
\end{align}
with integer $M,N,L,P,Q,R > 0$. 

It is now possible to numerically investigate many different momenta configurations. 
In order to cover two characteristic 
cases, namely $ k_A \sim k_B $ and $ k_A < k_B $ (or equivalently $k_A > k_B$)
 we choose the following two fixed configurations for  the transverse momenta 
of the forward jets: $\left( k_A, k_B \right)$ =  $(40, 50)$  
and $\left( k_A, k_B \right)$ = $(30, 60)$ GeV. 
We also fix the rapidities of the four tagged jets to the values 
$Y_A = 9$, $y_1 = 6$, $Y_2 = 3$, and $Y_B = 0$ 
whereas the two inner jets can have transverse momenta 
in the range $20 < k_{1,2} < 80$ GeV.

\begin{figure}[p]
\vspace{-2cm}
\centering
   \includegraphics[scale=0.60]{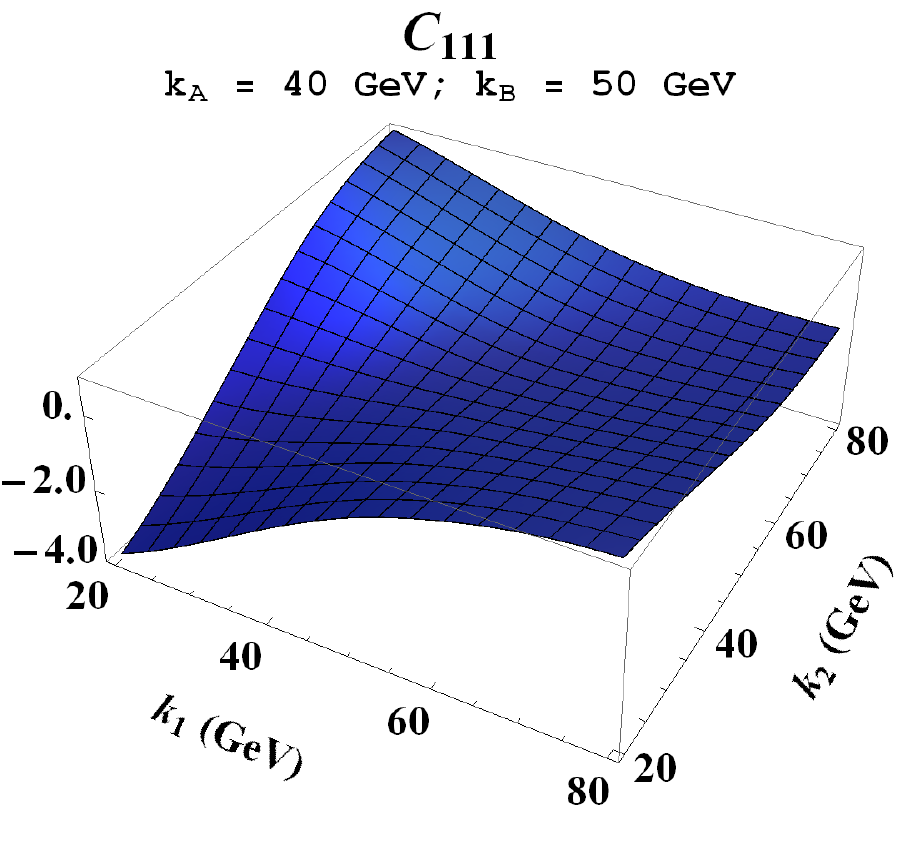}
   \includegraphics[scale=0.60]{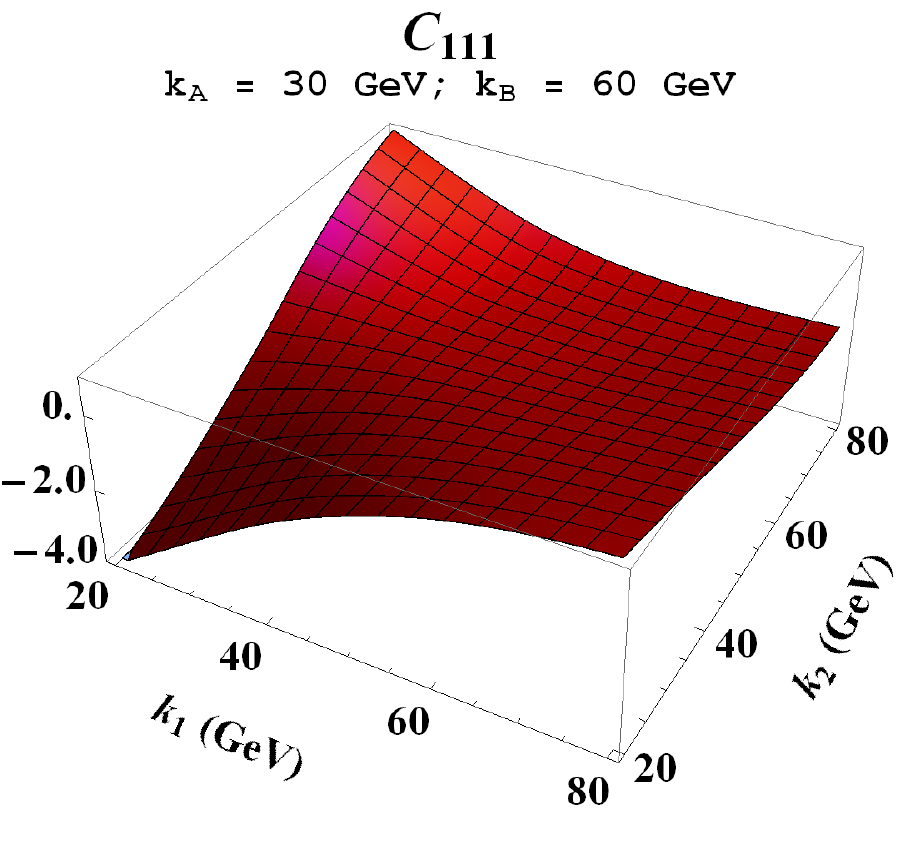}

   \includegraphics[scale=0.60]{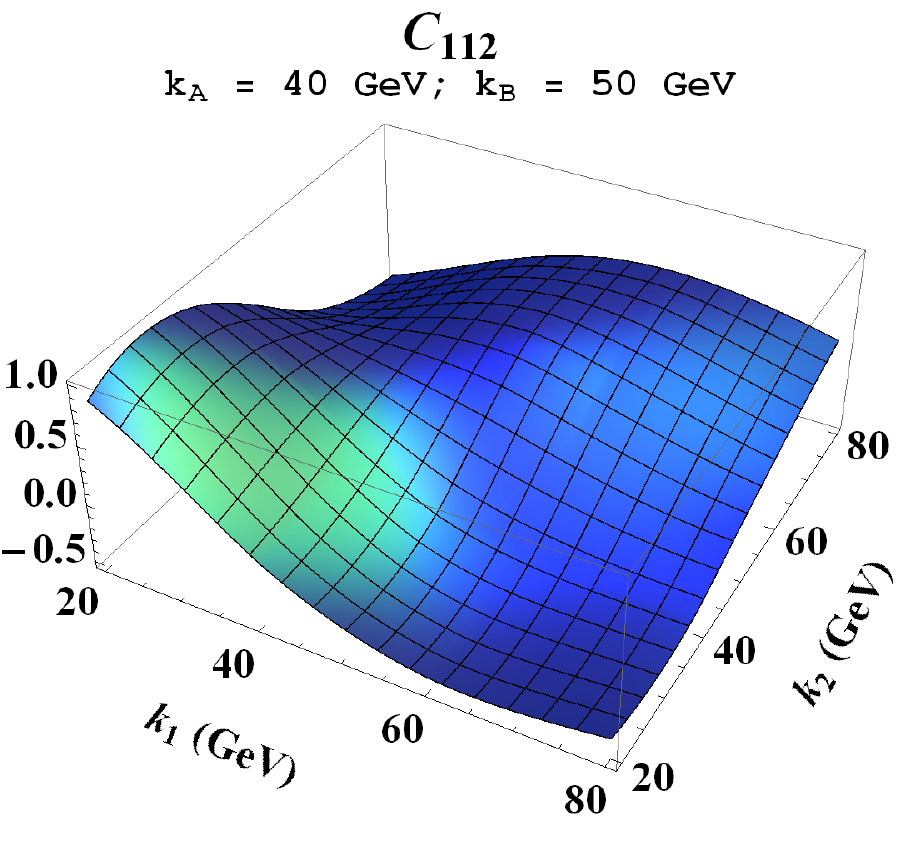}
   \includegraphics[scale=0.60]{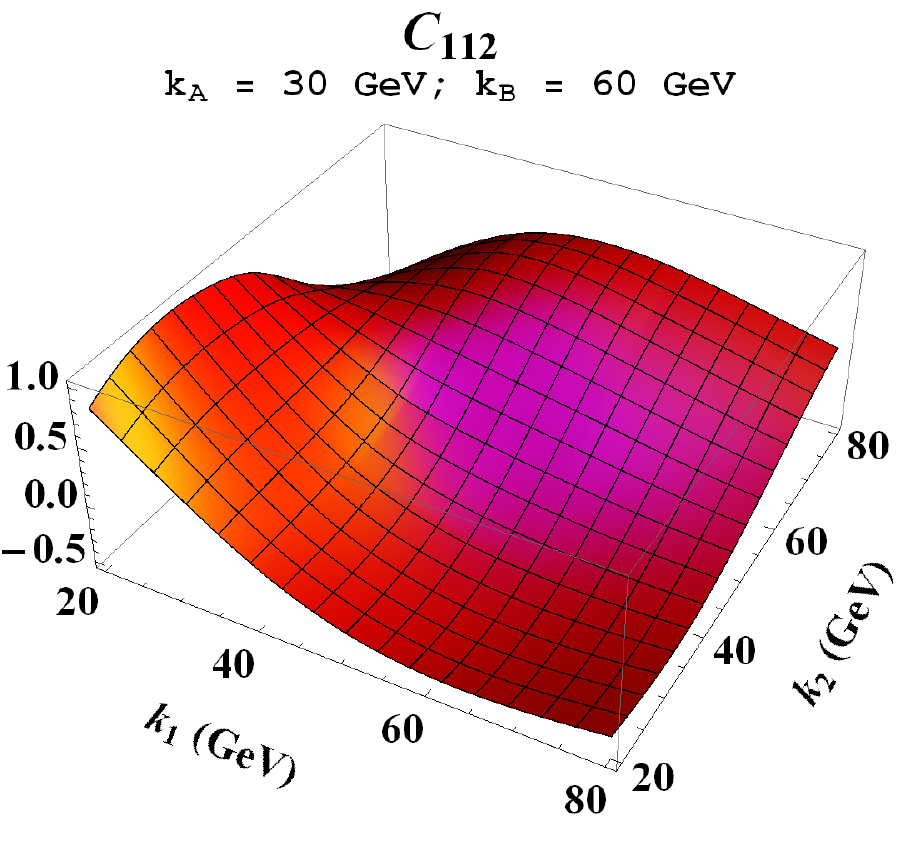}
   
   \includegraphics[scale=0.60]{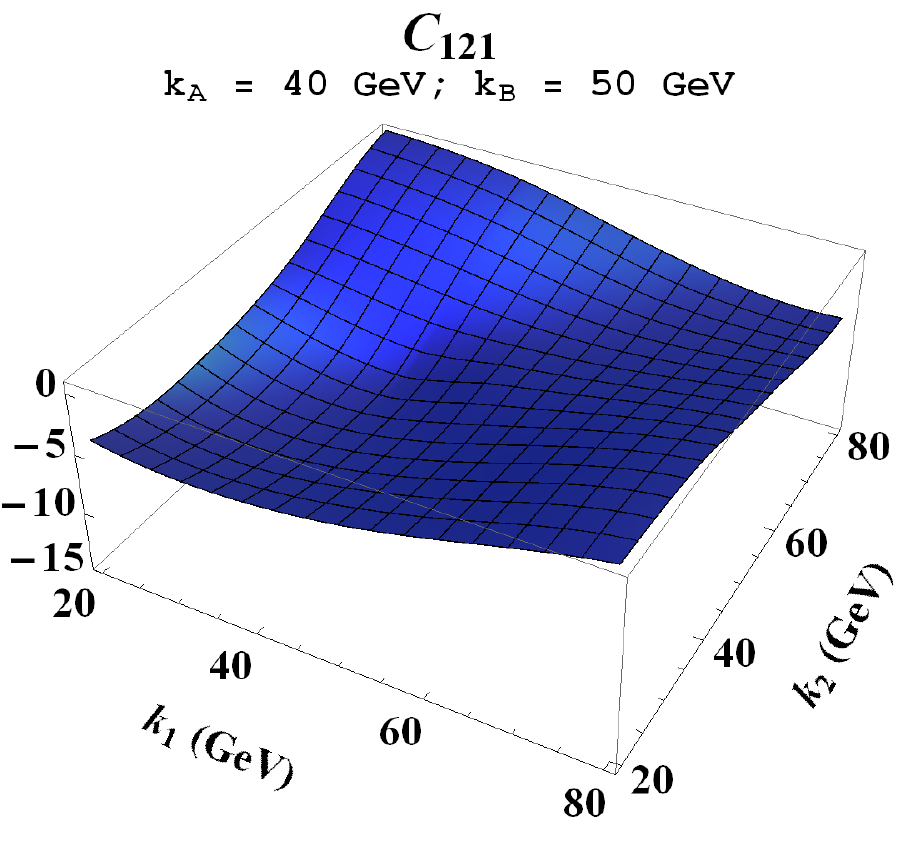}
   \includegraphics[scale=0.60]{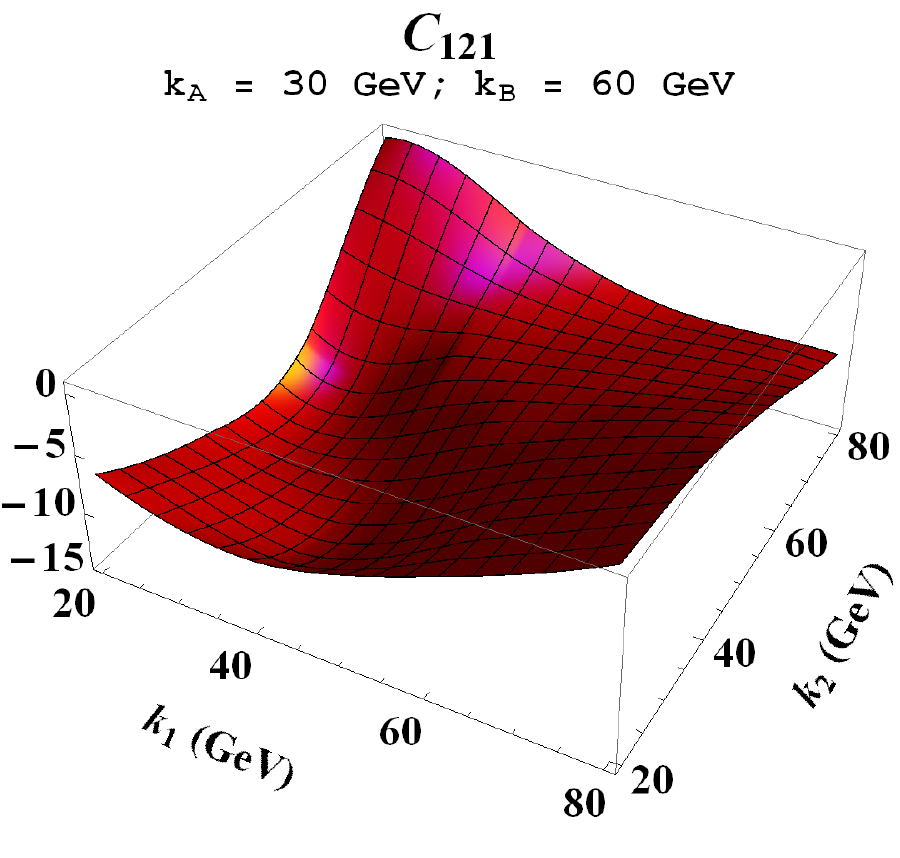}
   
   \includegraphics[scale=0.60]{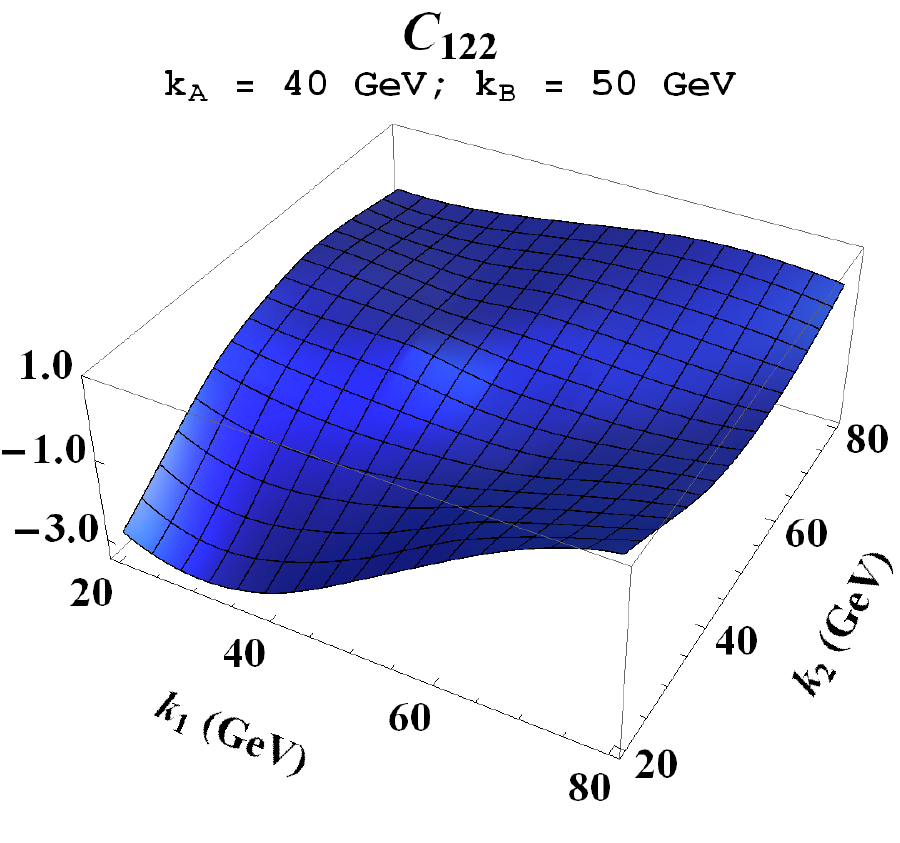}
   \includegraphics[scale=0.60]{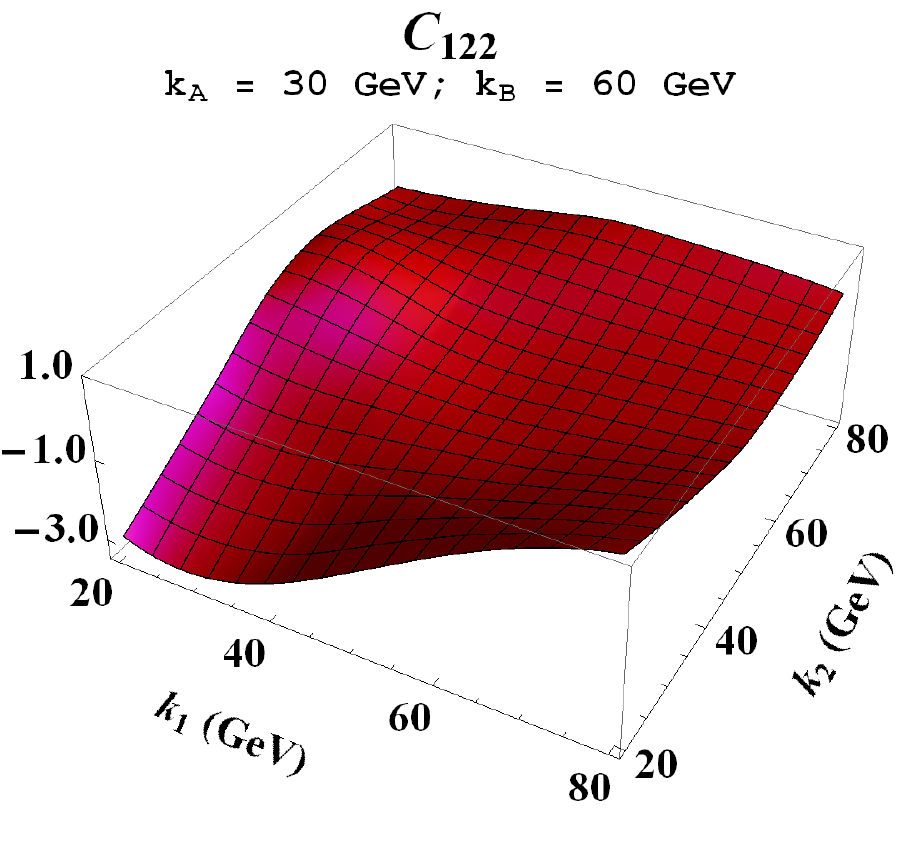}

\caption{\small $k_{1,2}$-dependence of the normalized ${\cal C}_{111}$, $C_{112}$, ${\cal C}_{121}$
and ${\cal C}_{122}$ for the two selected cases of forward jet 
transverse momenta $k_A$ and $k_B$.}
\label{C1nl}
\end{figure}

\begin{figure}[p]
\vspace{-2.0cm}
\centering
   \includegraphics[scale=0.60]{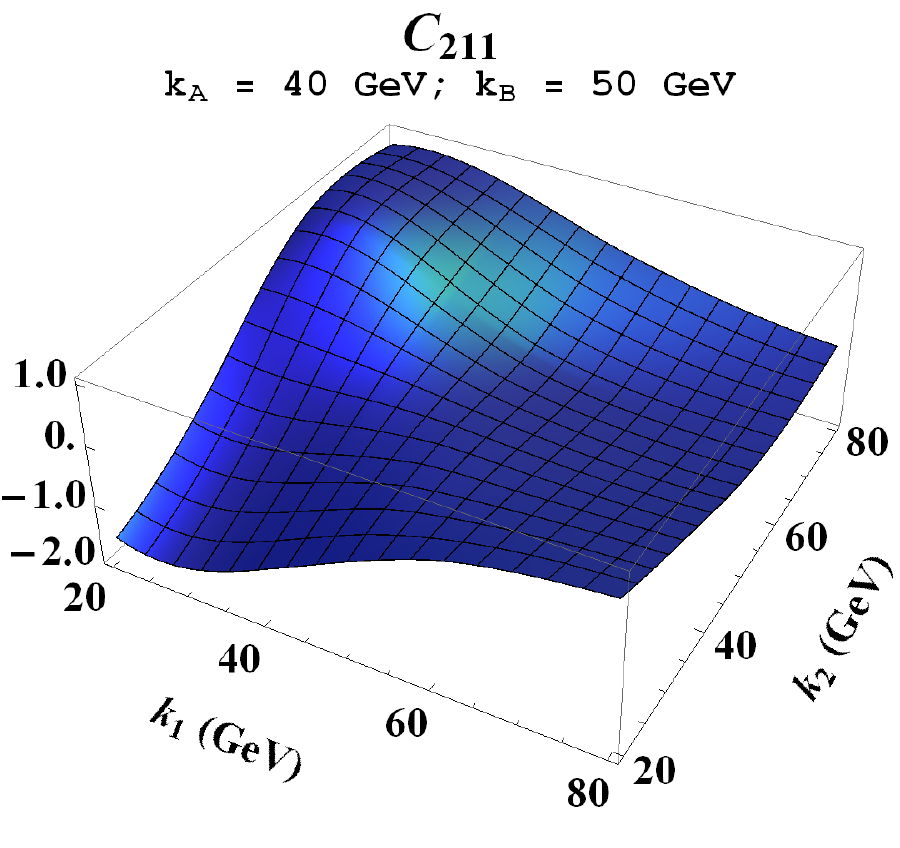}
   \includegraphics[scale=0.60]{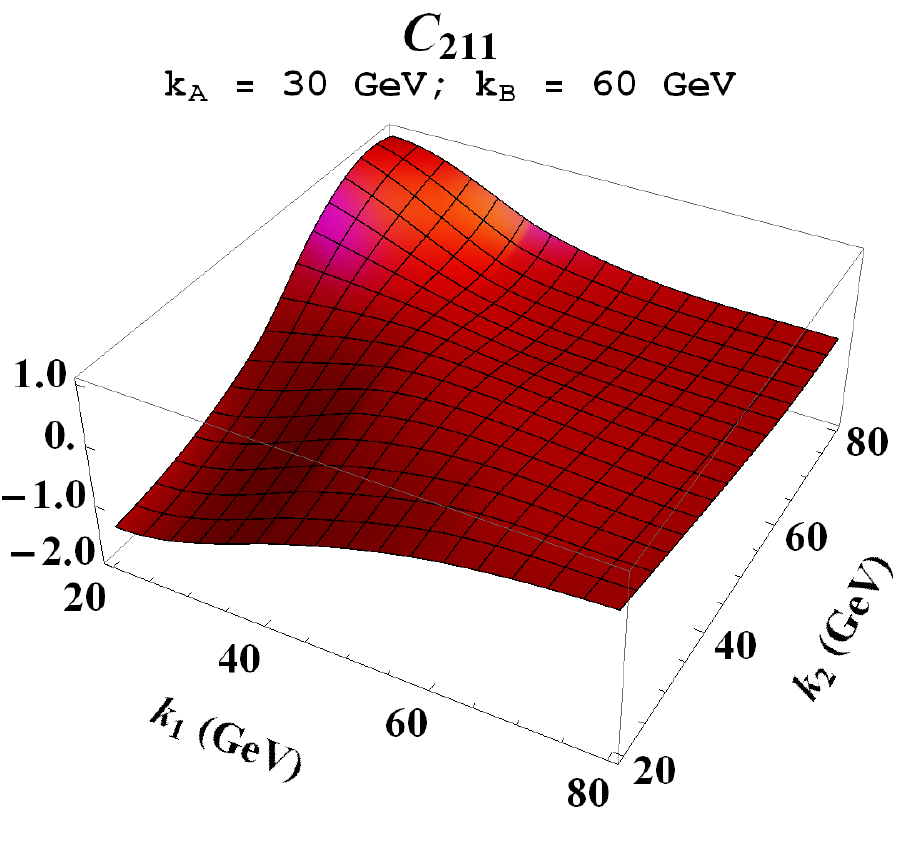}

   \includegraphics[scale=0.60]{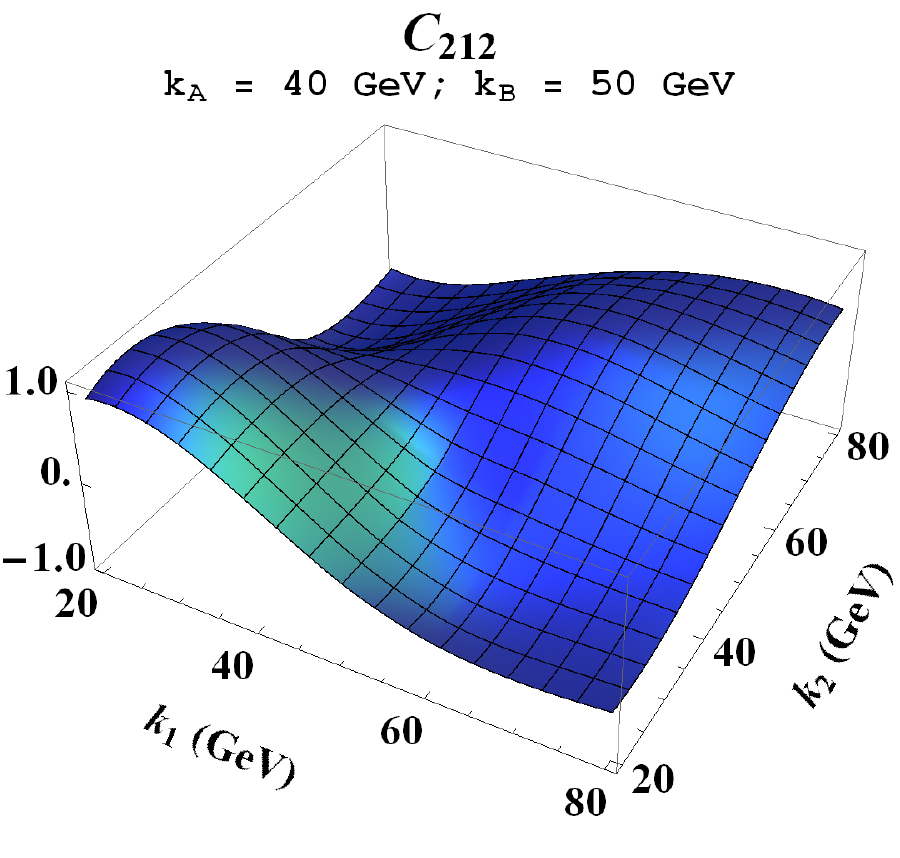}
   \includegraphics[scale=0.60]{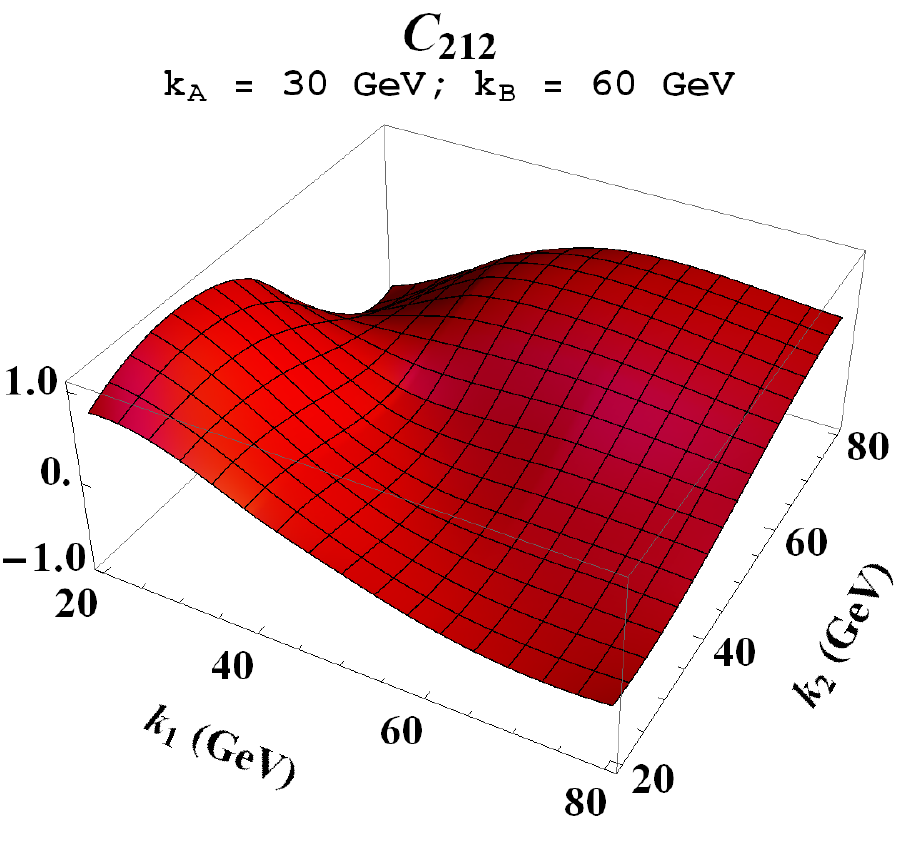}
   
   \includegraphics[scale=0.60]{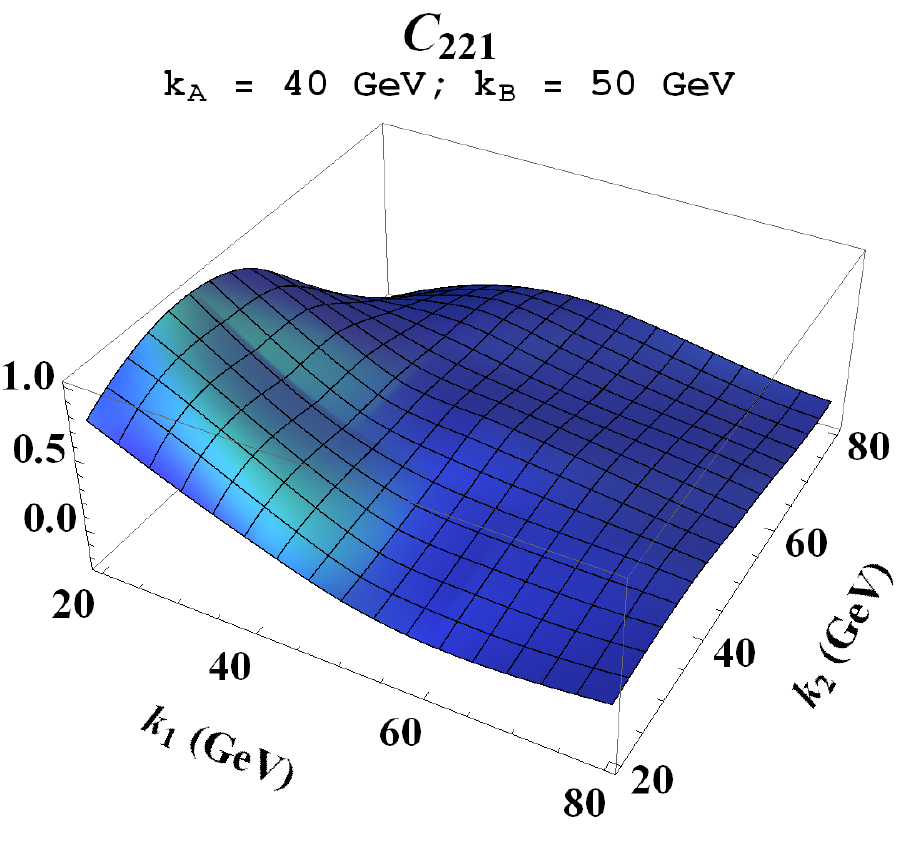}
   \includegraphics[scale=0.60]{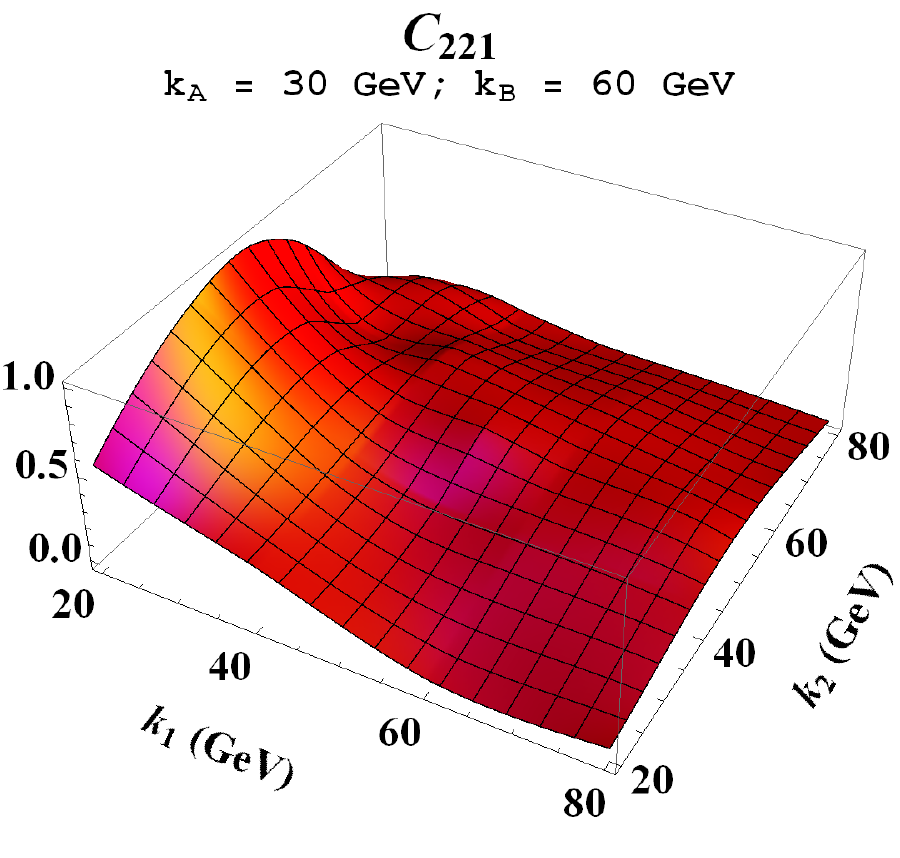}
   
   \includegraphics[scale=0.60]{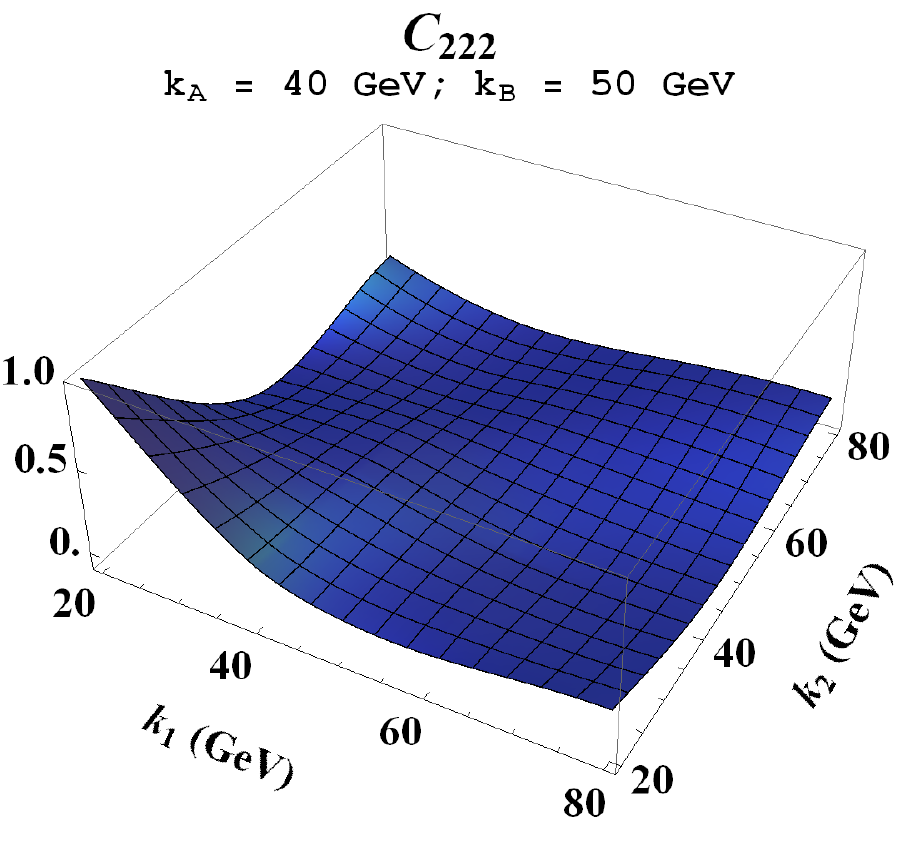}
   \includegraphics[scale=0.60]{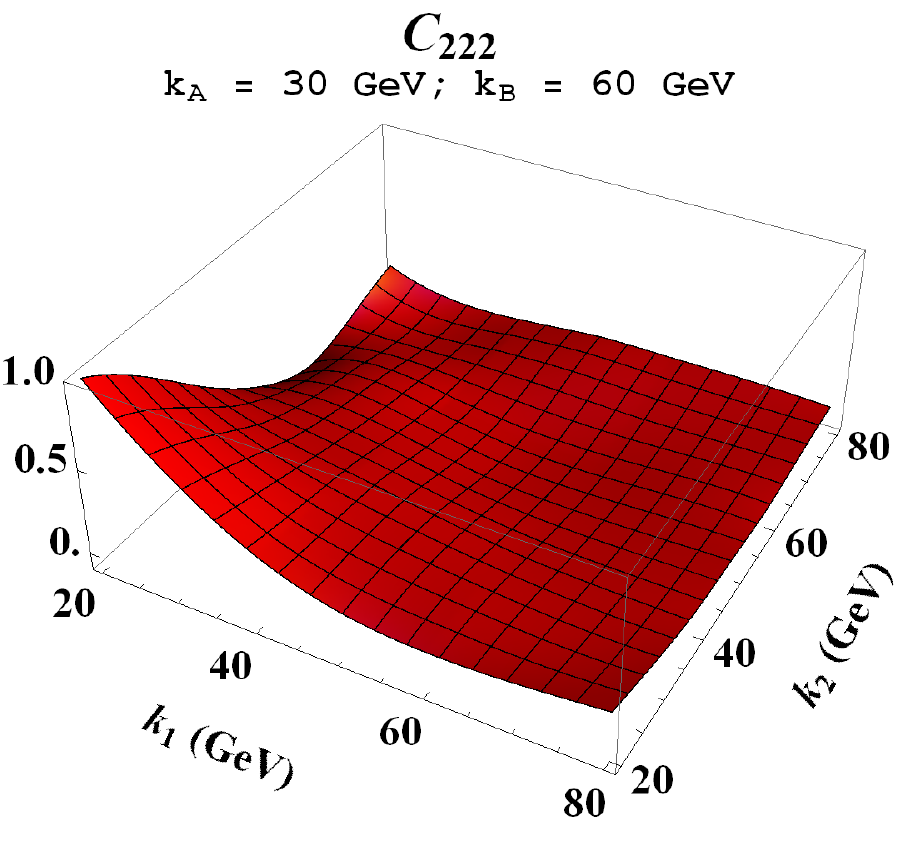}

\caption{\small $k_{1,2}$-dependence of the normalized ${\cal C}_{211}$, ${\cal C}_{212}$, ${\cal C}_{221}$
and ${\cal C}_{222}$ for the two selected cases of forward jet 
transverse momenta $k_A$ and $k_B$.}
\label{C2nl}
\end{figure}

In Fig.~\ref{C1nl} we present our results for the normalized
coefficients ${\cal C}_{111}$, ${\cal C}_{112}$, 
${\cal C}_{121}$ and ${\cal C}_{122}$ after they are divided by their respective maximum.
We find that the distributions are quite similar for the two configurations here chosen ($\left( k_A, k_B \right)$ = $(40, 50)$, $(30, 60)$ GeV) apart from the coefficient ${\cal C}_{121}$ which is quite more negative for the latter configuration when the transverse momentum of the first central jet, $k_1$, is low. 
Further coefficients, normalized as above, 
are calculated in Fig.~\ref{C2nl} for the cases ${\cal C}_{211}$, ${\cal C}_{212}$, ${\cal C}_{221}$
and ${\cal C}_{222}$. Again they are rather similar with the exception of  ${\cal C}_{221}$ at low $p_t$ of one of the centrals jets with largest rapidity. 

\begin{figure}[p]
\vspace{-2.0cm}
\centering
   \includegraphics[scale=0.65]{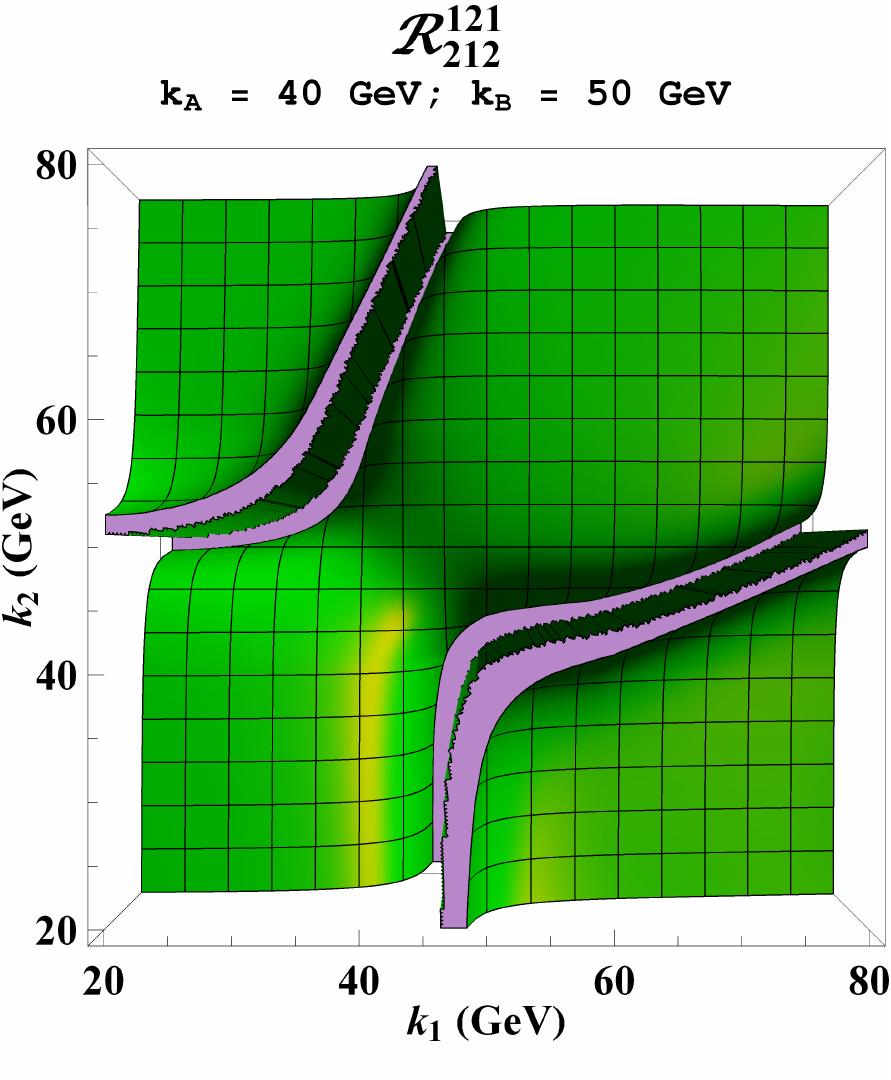}
   \includegraphics[scale=0.65]{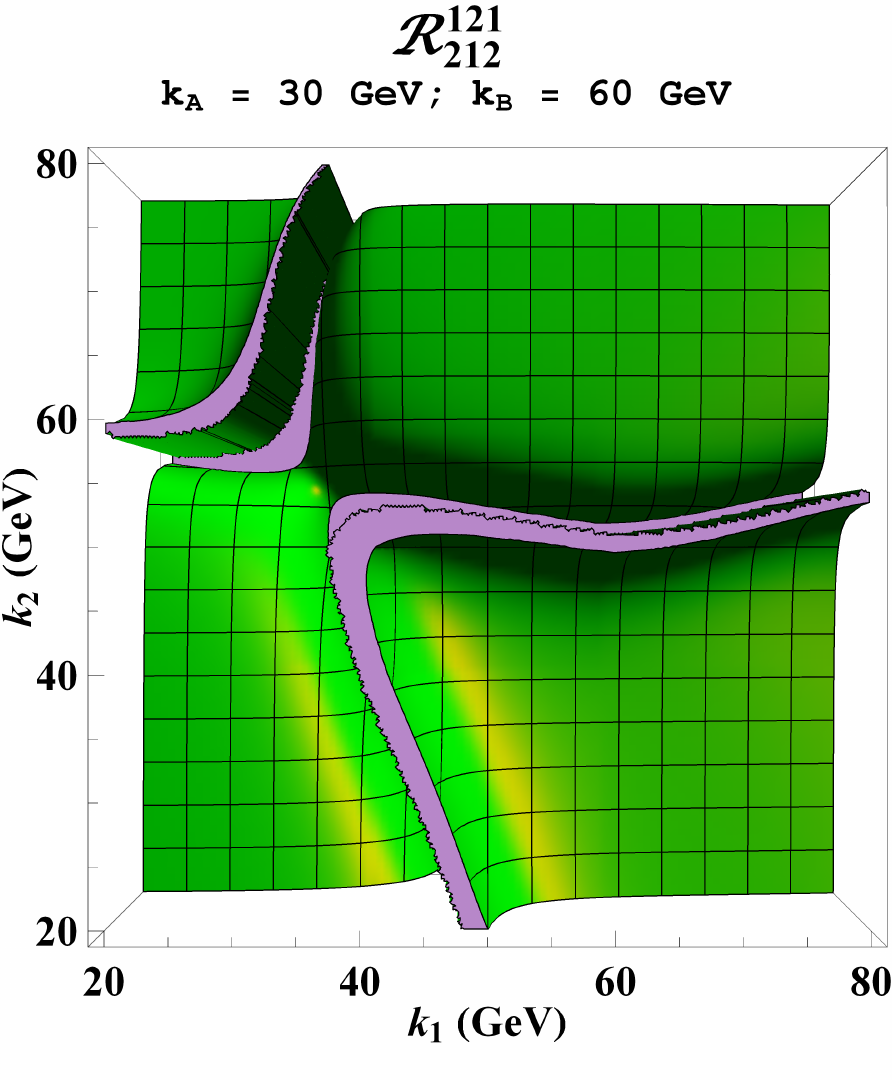}

   \includegraphics[scale=0.65]{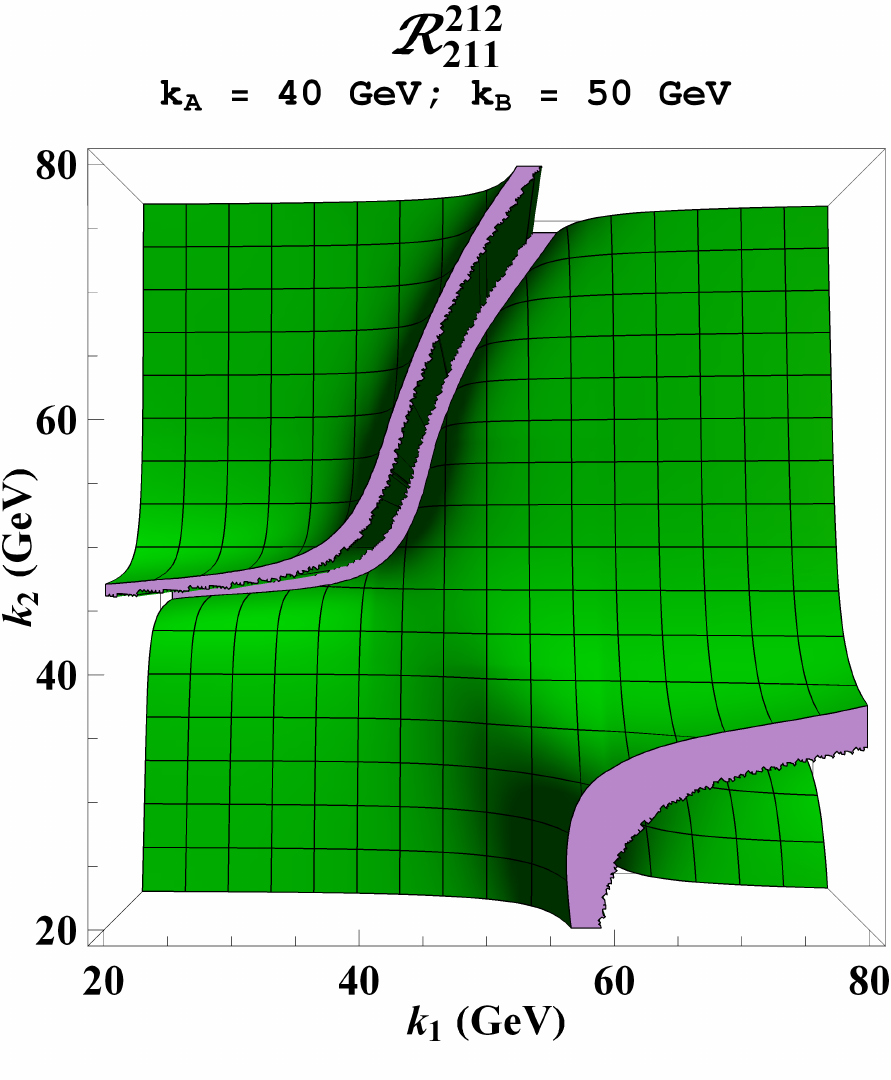}
   \includegraphics[scale=0.65]{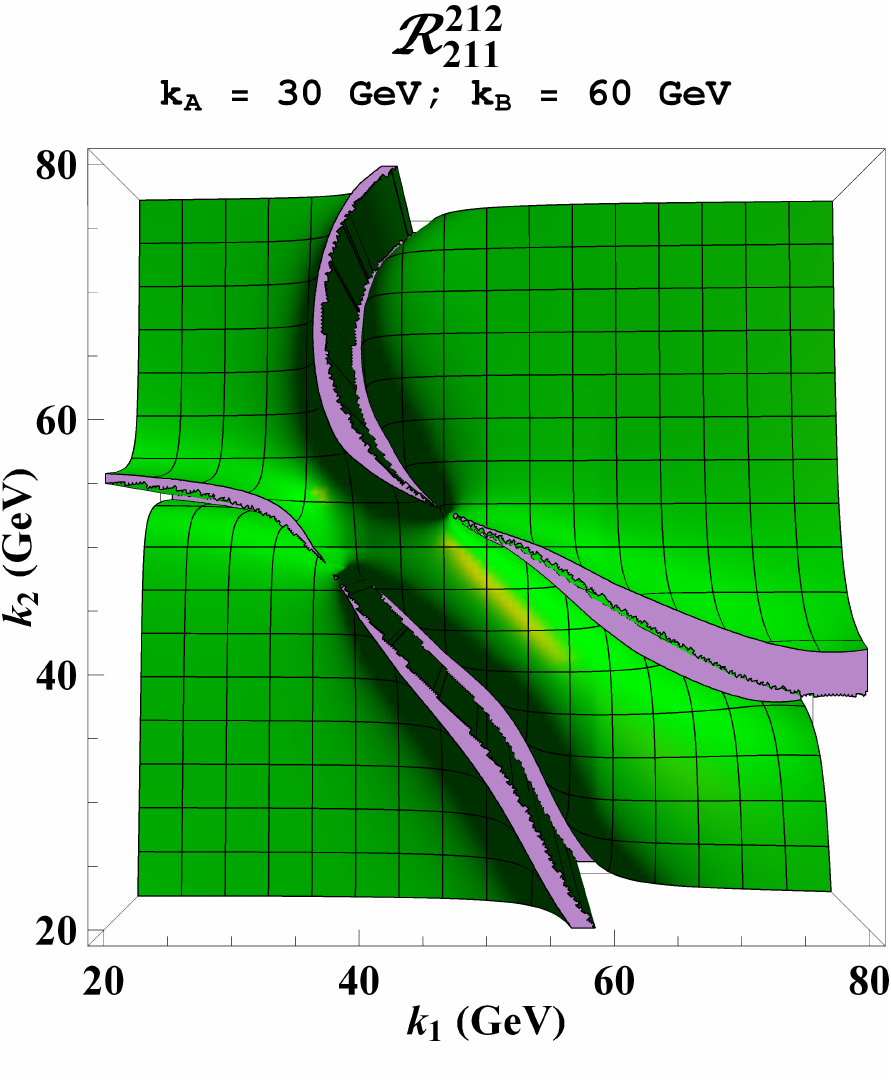}
   
   \includegraphics[scale=0.65]{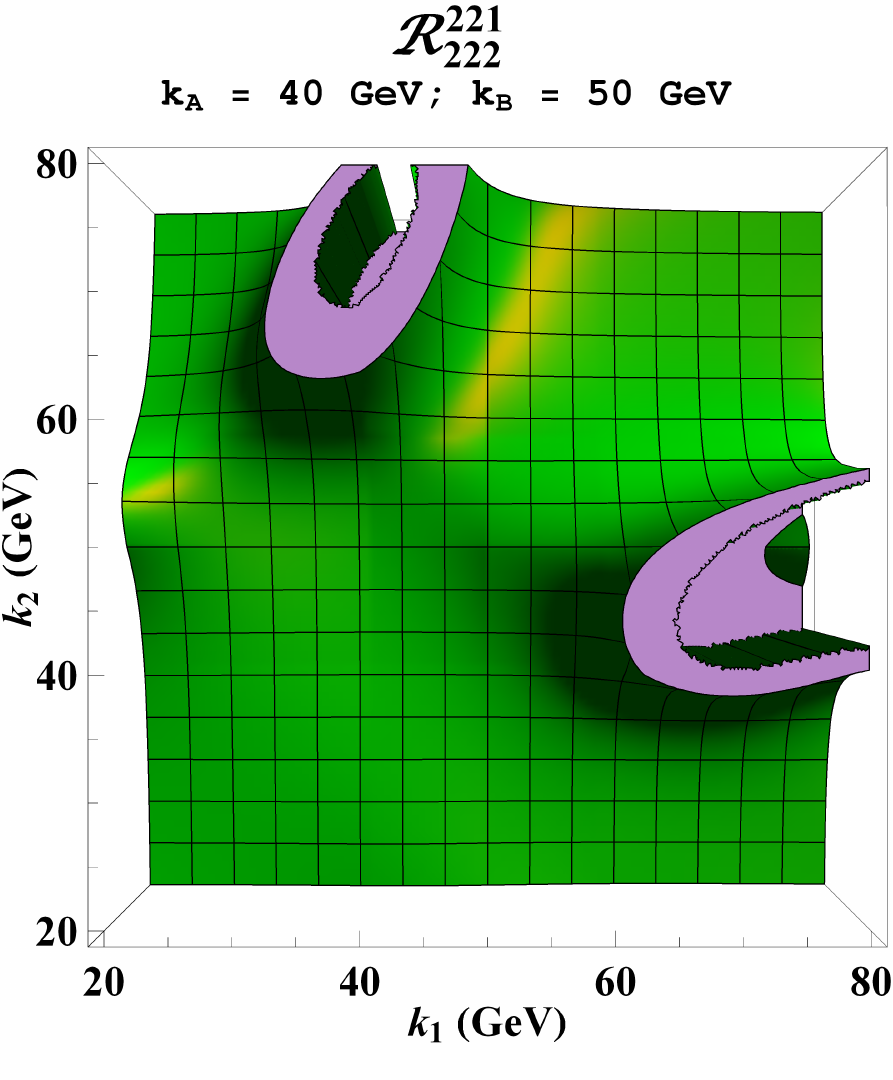}
   \includegraphics[scale=0.65]{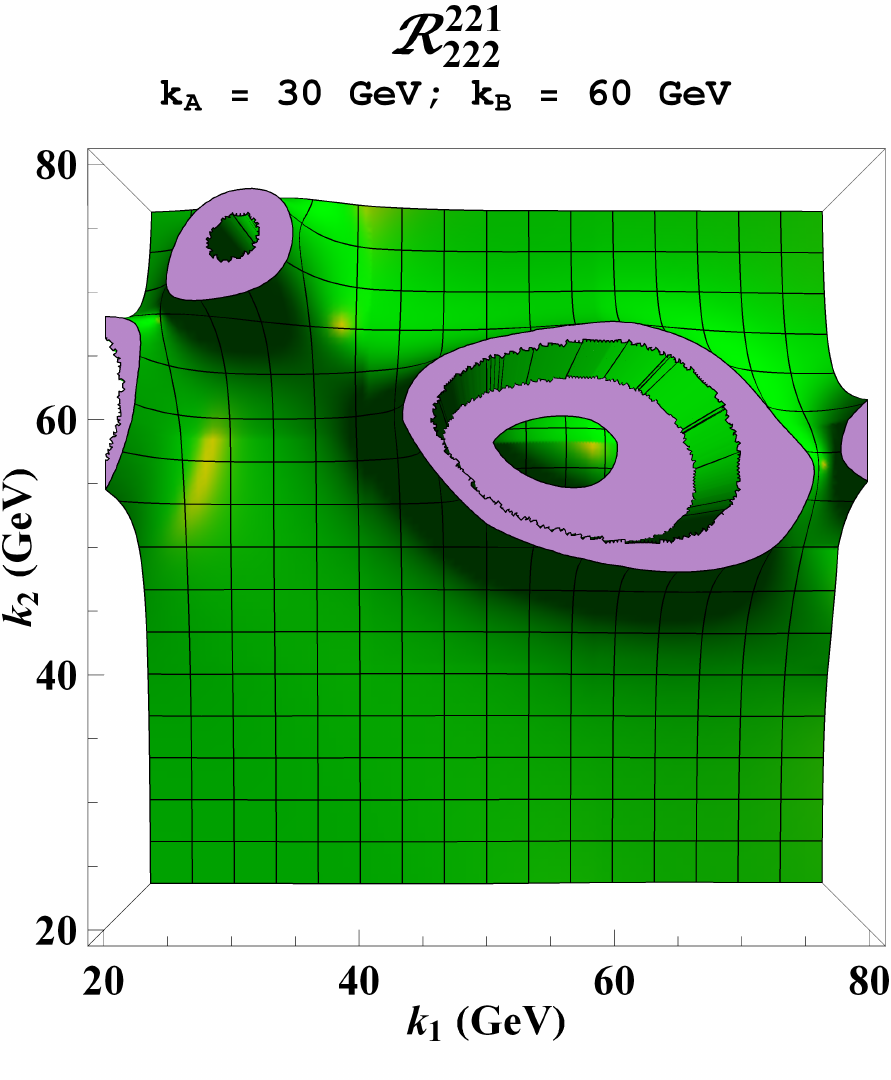}

\caption{\small $k_{1,2}$-dependence of $\mathcal{R}^{121}_{212}$, 
$\mathcal{R}^{212}_{211}$ and $\mathcal{R}^{221}_{222}$ 
for the two selected cases of forward/backward jets  
transverse momenta $k_A$ and $k_B$.}
\label{fig:ratios_1}
\end{figure}

\begin{figure}[p]
\vspace{-2.0cm}
\centering
   \includegraphics[scale=0.60]{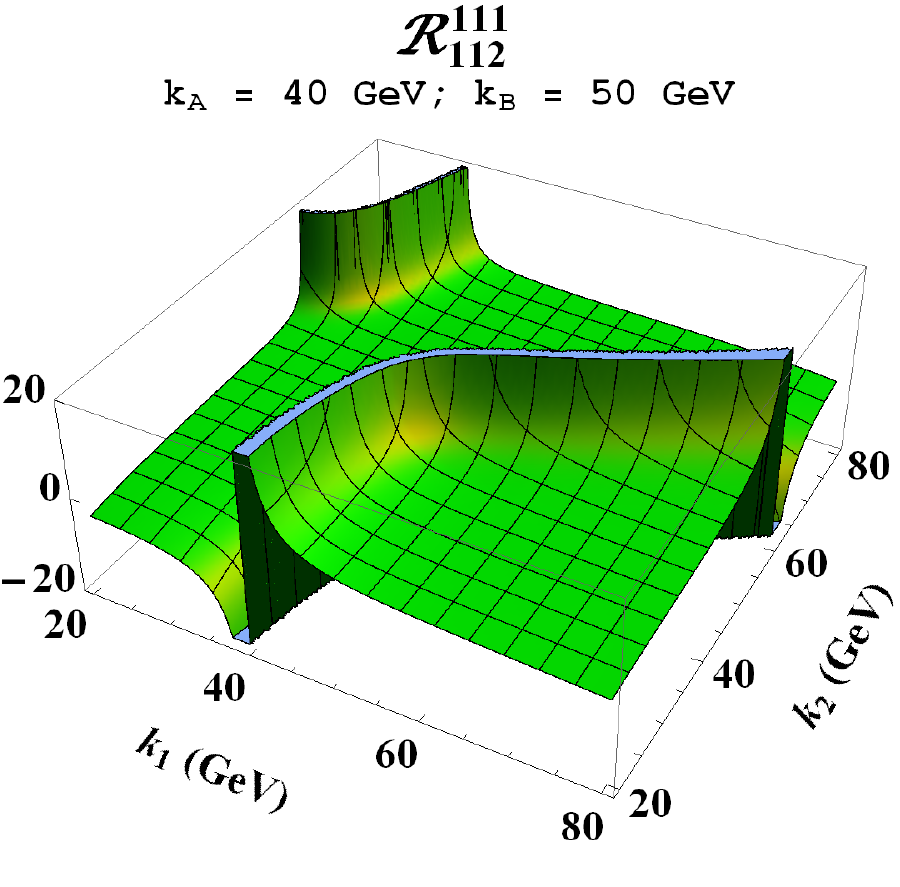}
   \includegraphics[scale=0.60]{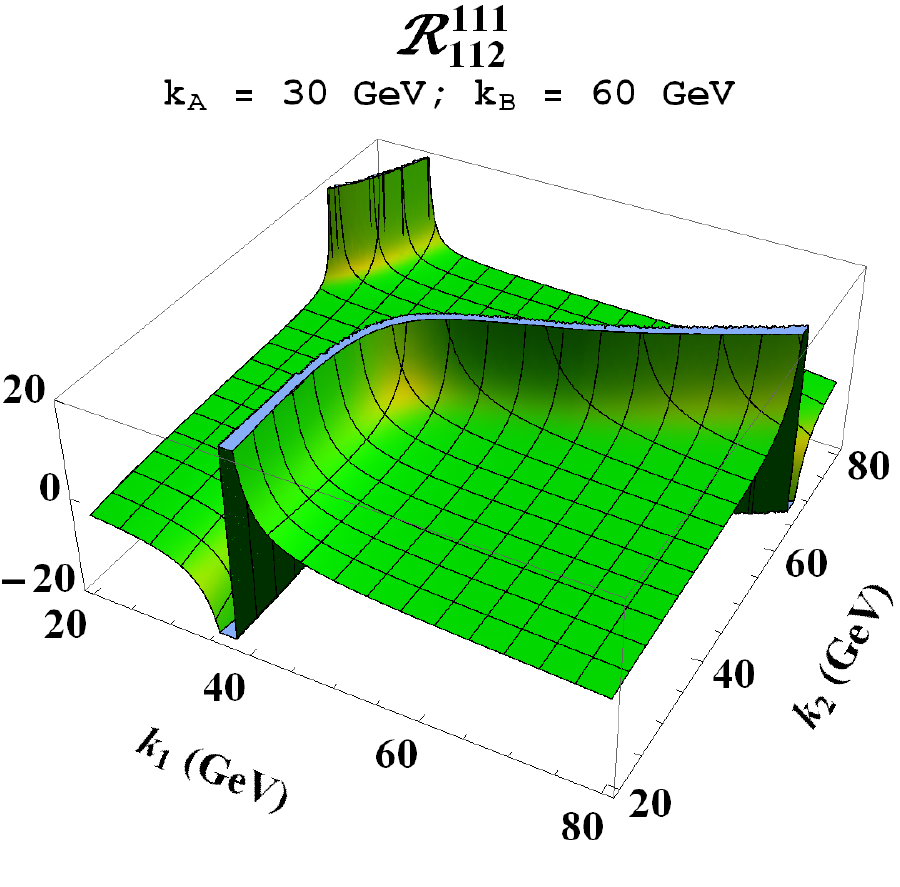}

   \includegraphics[scale=0.60]{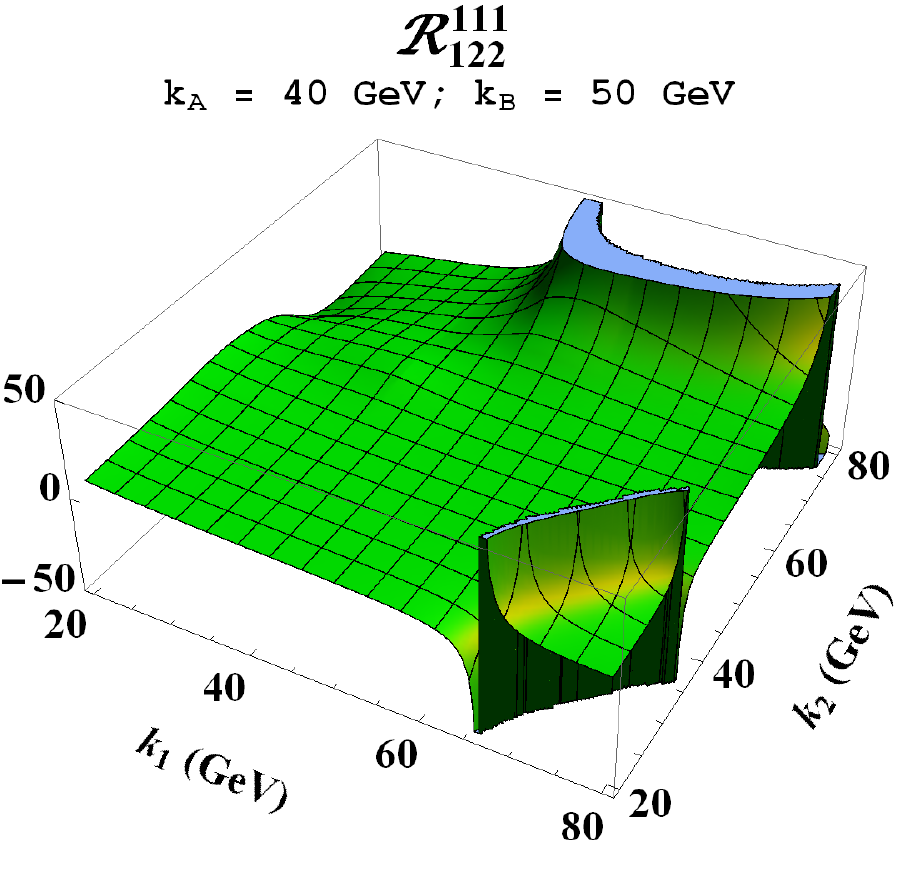}
   \includegraphics[scale=0.60]{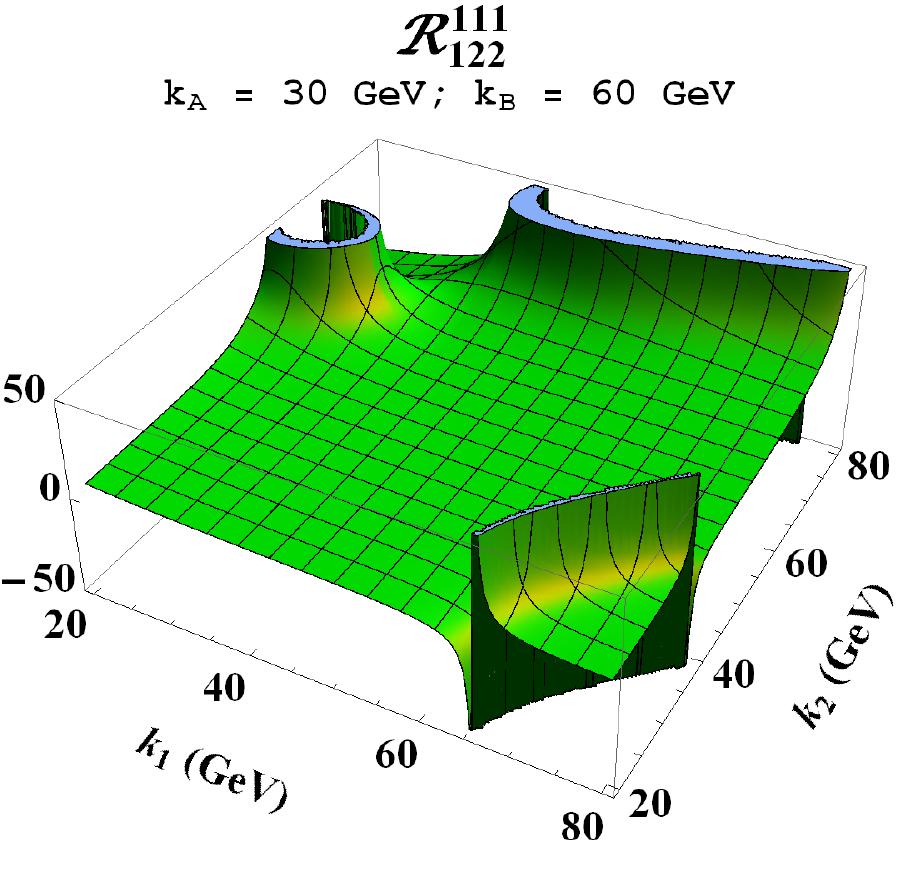}
   
   \includegraphics[scale=0.60]{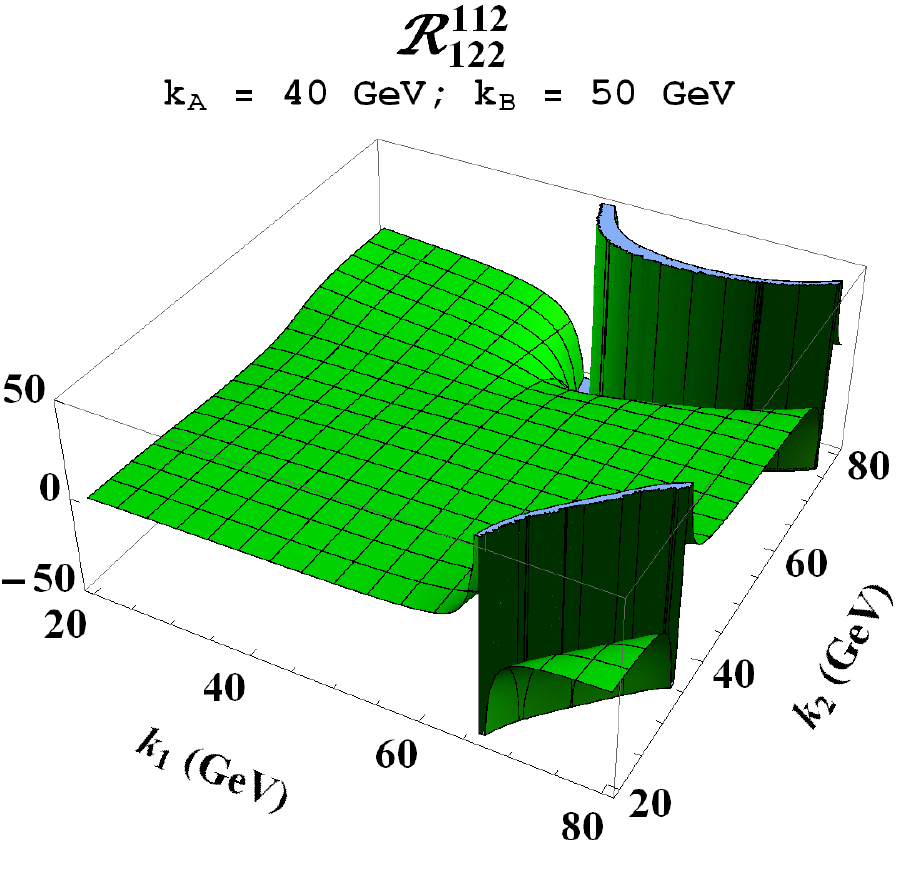}
   \includegraphics[scale=0.60]{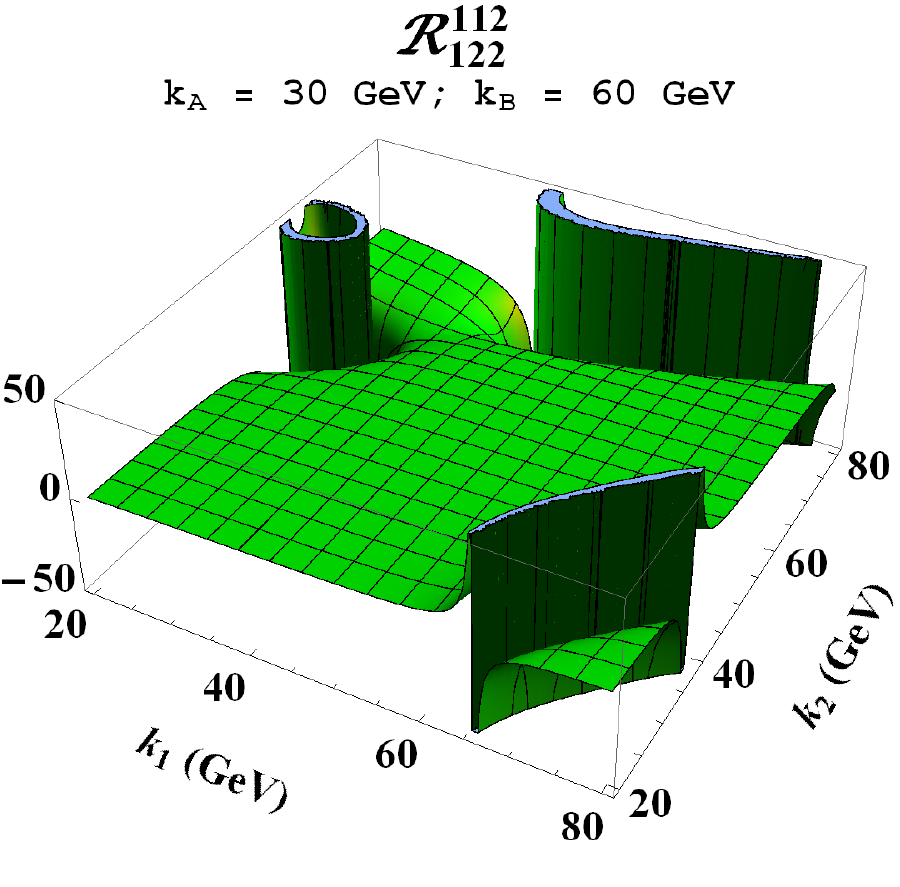}
   
   \includegraphics[scale=0.60]{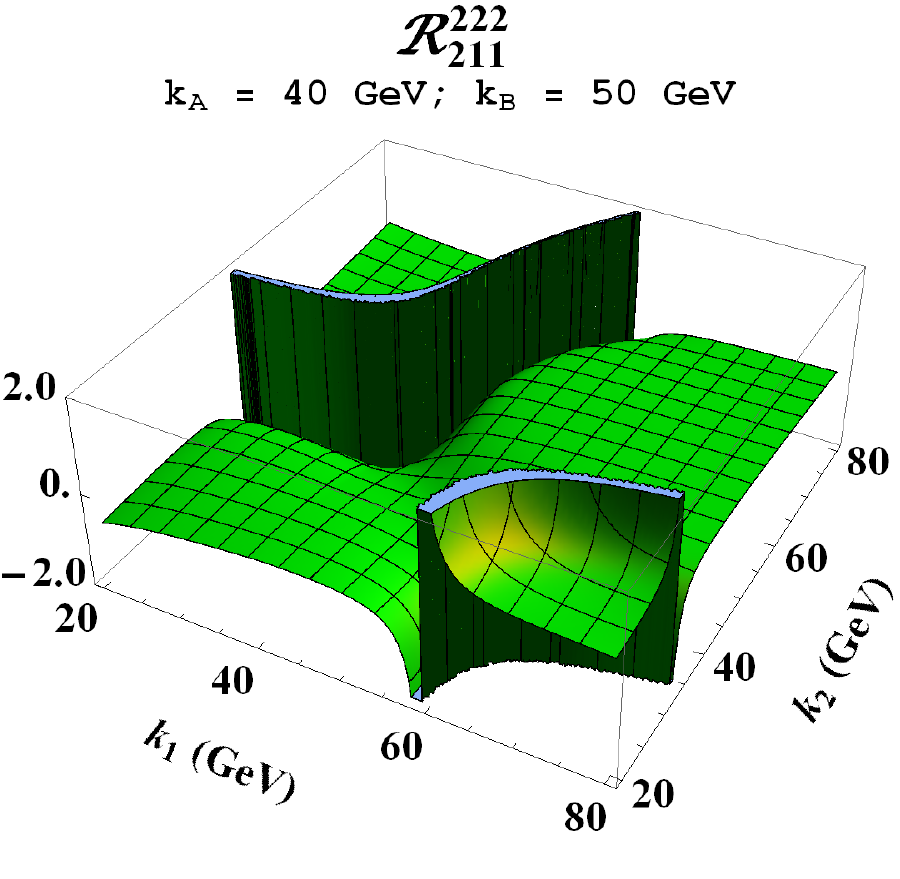}
   \includegraphics[scale=0.60]{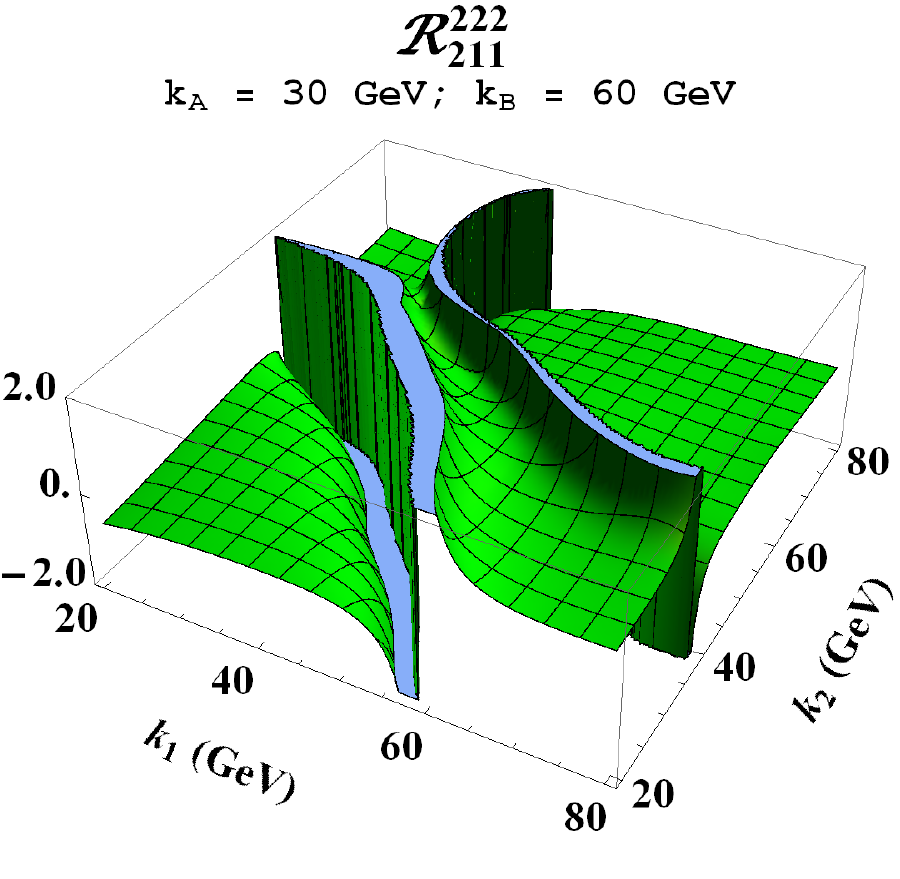}

\caption{\small $k_{1,2}$-dependence of 
$\mathcal{R}^{111}_{112}$, $\mathcal{R}^{111}_{122}$, 
$\mathcal{R}^{112}_{122}$ and $\mathcal{R}^{222}_{211}$ 
for the two selected cases of forward/backward jets  
transverse momenta $k_A$ and $k_B$.}
\label{fig:ratios_2}
\end{figure}

Since these coefficients change sign on the parameter space here studied, it is clear that for the associated ratios $\mathcal{R}^{MNL}_{PQR}$  there will be some lines of singularities. We have investigated $\mathcal{R}^{121}_{212}$,  $\mathcal{R}^{212}_{211}$ and $\mathcal{R}^{221}_{222}$ in Fig.~\ref{fig:ratios_1}. 
In this case the configurations $\left( k_A, k_B \right)$ = $(40, 50)$, $(30, 60)$ GeV behave quite differently. This is due to the variation of the position of the zeroes of those coefficients ${\cal C}_{MNP}$ 
chosen as denominators in these quantities. It would be very interesting to test if these singularity lines are present in any form in the LHC experimental data. A further set of ratios, $\mathcal{R}^{111}_{112}$, $\mathcal{R}^{111}_{122}$, $\mathcal{R}^{112}_{122}$ and $\mathcal{R}^{222}_{211}$ , with their characteristic singular lines, is presented in Fig.~\ref{fig:ratios_2}.
In general, we have found a very weak dependence on variations of the rapidity of the more central jets $y_{1,2}$ for all the observables here presented.  

We have used both  \textsc{Fortran} and \textsc{Mathematica} for
the numerical computation
of the ratios $\mathcal{R}^{MNL}_{PQR}$. 
We made extensive use of the integration routine \cod{Vegas}~\cite{VegasLepage:1978} 
as implemented in the \cod{Cuba} library~\cite{Cuba:2005,ConcCuba:2015}.
Furthermore, we used the \cod{Quadpack} library~\cite{Quadpack:book:1983}
and a slightly modified version of the \cod{Psi}~\cite{RpsiCody:1973} routine.

We conclude here our numerical analysis. We have offered several interesting observables where BFKL effects could have sizable effects. More detailed calculations are needed, including the introduction of parton distribution functions effects and higher order terms in jet vertices and Green functions. However, we argue that the bulk of the relevant contributions are already contained in the calculations here discussed, in particular, for the ratios $\mathcal{R}^{MNL}_{PQR}$. It will be very important to 
compare against the BFKL Monte Carlo code \cod{BFKLex}~\cite{Chachamis:2013rca,Caporale:2013bva,Chachamis:2012qw,Chachamis:2012fk,Chachamis:2011nz,Chachamis:2011rw,Chachamis:2015zzp}
as well as to calculate the same quantities
with other, more conventional, approaches~\cite{Bury:2015dla,Nefedov:2013ywa,vanHameren:2012uj}
 in order to gauge if they differ from our results. This includes those analysis where the four-jet predictions stem from two independent gluon ladders~\cite{Maciula:2015,Maciula:2014pla}.

\section{Summary \& Outlook}

We have presented new observables to study four-jet production at hadron colliders in terms of its azimuthal angle dependences. These correspond to the ratios of correlation functions of products of cosines of azimuthal angle differences among the tagged jets. We used a single BFKL ladder approach,  with inclusive production of two forward/backward and two further, more central, tagged jets. The dependence on the transverse momenta and rapidities of the two central jets is a distinct signal of BFKL dynamics. For future works, more accurate analysis are needed: introduction of parton distribution functions, higher order effect and realistic experimental cuts. It is also pressing to calculate our proposed observables using other approaches not based on the BFKL approach and to test how they can differ from our predictions. Last but not least, we encourage our experimental colleagues to analyze these observables in recent and future LHC data.

\begin{flushleft}
{\bf \large Acknowledgements}
\end{flushleft}
G.C. acknowledges support from the MICINN, Spain, 
under contract FPA2013-44773-P. 
A.S.V. acknowledges support from Spanish Government 
(MICINN (FPA2010-17747,FPA2012-32828)) and, together with F.C. and F.G.C., 
to the Spanish MINECO Centro de Excelencia Severo Ochoa Programme 
(SEV-2012-0249). 
F.G.C. thanks the Instituto de Fisica Teorica 
(IFT UAM-CSIC) in Madrid for warm hospitality.


\begin{thebibliography}{10}

\bibitem{Lipatov:1985uk}
  L.~N.~Lipatov,
  Sov.\ Phys.\ JETP {\bf 63} (1986) 904
   [Zh.\ Eksp.\ Teor.\ Fiz.\  {\bf 90} (1986) 1536].
  
\bibitem{Balitsky:1978ic}
  I.~I.~Balitsky and L.~N.~Lipatov,
  Sov.\ J.\ Nucl.\ Phys.\  {\bf 28} (1978) 822
   [Yad.\ Fiz.\  {\bf 28} (1978) 1597].
  
\bibitem{Kuraev:1977fs}
  E.~A.~Kuraev, L.~N.~Lipatov and V.~S.~Fadin,
  Sov.\ Phys.\ JETP {\bf 45} (1977) 199
   [Zh.\ Eksp.\ Teor.\ Fiz.\  {\bf 72} (1977) 377].
  
\bibitem{Kuraev:1976ge}
  E.~A.~Kuraev, L.~N.~Lipatov and V.~S.~Fadin,
  Sov.\ Phys.\ JETP {\bf 44} (1976) 443
   [Zh.\ Eksp.\ Teor.\ Fiz.\  {\bf 71} (1976) 840]
   [Erratum-ibid.\  {\bf 45} (1977) 199].
  
\bibitem{Lipatov:1976zz}
  L.~N.~Lipatov,
  Sov.\ J.\ Nucl.\ Phys.\  {\bf 23} (1976) 338
   [Yad.\ Fiz.\  {\bf 23} (1976) 642].
  
\bibitem{Fadin:1975cb}
  V.~S.~Fadin, E.~A.~Kuraev and L.~N.~Lipatov,
  Phys.\ Lett.\ B {\bf 60} (1975) 50.
  
\bibitem{Fadin:1998py}
  V.~S.~Fadin and L.~N.~Lipatov,
  Phys.\ Lett.\ B {\bf 429} (1998) 127
  [hep-ph/9802290].
  
\bibitem{Ciafaloni:1998gs}
  M.~Ciafaloni and G.~Camici,
  Phys.\ Lett.\ B {\bf 430} (1998) 349
  [hep-ph/9803389].

\bibitem{Hentschinski:2012kr}
  M.~Hentschinski, A.~Sabio Vera and C.~Salas,
  Phys.\ Rev.\ Lett.\  {\bf 110} (2013) 041601
  [arXiv:1209.1353 [hep-ph]].
  
\bibitem{Hentschinski:2013id}
  M.~Hentschinski, A.~Sabio Vera and C.~Salas,
  Phys.\ Rev.\ D {\bf 87} (2013) 076005
  [arXiv:1301.5283 [hep-ph]].
   
\bibitem{Mueller:1986ey}
  A.~H.~Mueller and H.~Navelet,
  Nucl.\ Phys.\ B {\bf 282} (1987) 727.
   
\bibitem{DelDuca:1993mn}
  V.~Del Duca and C.~R.~Schmidt,
  Phys.\ Rev.\ D {\bf 49} (1994) 4510
  [hep-ph/9311290].
  
\bibitem{Stirling:1994he}
  W.~J.~Stirling,
  Nucl.\ Phys.\ B {\bf 423} (1994) 56
  [hep-ph/9401266].
  
\bibitem{Orr:1997im}
  L.~H.~Orr and W.~J.~Stirling,
  Phys.\ Rev.\ D {\bf 56} (1997) 5875
  [hep-ph/9706529].
  
\bibitem{Kwiecinski:2001nh}
  J.~Kwiecinski, A.~D.~Martin, L.~Motyka and J.~Outhwaite,
  Phys.\ Lett.\ B {\bf 514} (2001) 355
  [hep-ph/0105039].
  
\bibitem{Vera:2006un}
  A.~Sabio Vera,
  Nucl.\ Phys.\ B {\bf 746} (2006) 1
  [hep-ph/0602250].
  
\bibitem{Vera:2007kn}
  A.~Sabio Vera and F.~Schwennsen,
  Nucl.\ Phys.\ B {\bf 776} (2007) 170
  [hep-ph/0702158 [HEP-PH]].
     
\bibitem{Ducloue:2013bva}
  B.~Ducloue, L.~Szymanowski and S.~Wallon,
  Phys.\ Rev.\ Lett.\  {\bf 112} (2014) 082003
  [arXiv:1309.3229 [hep-ph]].

\bibitem{Caporale:2014gpa}
  F.~Caporale, D.~Y.~Ivanov, B.~Murdaca and A.~Papa,
  Eur.\ Phys.\ J.\ C {\bf 74} (2014) 3084
  [arXiv:1407.8431 [hep-ph]].
  
\bibitem{Celiberto:2015dgl}
 F.~G. Celiberto, D.~Yu. Ivanov, B.~Murdaca and A.~Papa,
 Eur.\ Phys.\ J.\ C {\bf 75} (2015) 292 
 [arXiv:1504.08233 [hep-ph]].
 
\bibitem{Ciesielski:2014dfa}
  R.~Ciesielski,
  arXiv:1409.5473 [hep-ex].
 
\bibitem{Angioni:2011wj} 
  M.~Angioni, G.~Chachamis, J.~D.~Madrigal and A.~Sabio Vera,
  Phys.\ Rev.\ Lett.\  {\bf 107}, 191601 (2011)
  [arXiv:1106.6172 [hep-th]].
 
 
\bibitem{Caporale:2015vya} 
  F.~Caporale, G.~Chachamis, B.~Murdaca and A.~Sabio Vera,
  arXiv:1508.07711 [hep-ph].
  To appear in Phys.\ Rev.\ Lett.\ 
   
\bibitem{Caporale:2013uva}
  F.~Caporale, B.~Murdaca, A.~Sabio Vera and C.~Salas,
  Nucl.\ Phys.\ B {\bf 875} (2013) 134
  [arXiv:1305.4620 [hep-ph]].
 
   
\bibitem{VegasLepage:1978}
 G.P.~Lepage, 
 J.\ Comput.\ Phys. {\bf 27} (1978) 192.

  
\bibitem{Cuba:2005}
 T.~Hahn,
  Comput.\ Phys.\ Commun. {\bf 168} (2005) 78
  [arXiv:1408.6373 [hep-ph]].
  
\bibitem{ConcCuba:2015}
 T.~Hahn,
  J.\ Phys.\ Conf.\ Ser. {\bf 608} (2015) 1
  [arXiv:hep-ph/0404043].
  
 
\bibitem{Quadpack:book:1983}
R.~Piessens, E.~De~Doncker-Kapenga and C.~W.~Überhuber,
 Springer, ISBN: 3-540-12553-1, 1983.

\bibitem{RpsiCody:1973}
 W.~J. Cody, A.~J. Strecok and H.~C. Thacher, 
 Math.\ Comput.\ {\bf 27} (1973) 121.
 


\bibitem{Chachamis:2013rca}
  G.~Chachamis and A.~Sabio Vera,
  PoS DIS {\bf 2013} (2013) 167
  [arXiv:1307.7750].
  
\bibitem{Caporale:2013bva}
  F.~Caporale, G.~Chachamis, J.~D.~Madrigal, B.~Murdaca and A.~Sabio Vera,
  Phys.\ Lett.\ B {\bf 724} (2013) 127
  [arXiv:1305.1474 [hep-th]].
  
\bibitem{Chachamis:2012qw}
  G.~Chachamis, A.~Sabio Vera and C.~Salas,
  Phys.\ Rev.\ D {\bf 87} (2013) 1,  016007
  [arXiv:1211.6332 [hep-ph]].
  
\bibitem{Chachamis:2012fk}
  G.~Chachamis and A.~Sabio Vera,
  Phys.\ Lett.\ B {\bf 717} (2012) 458
  [arXiv:1206.3140 [hep-th]].
  
\bibitem{Chachamis:2011nz}
  G.~Chachamis and A.~Sabio Vera,
  Phys.\ Lett.\ B {\bf 709} (2012) 301
  [arXiv:1112.4162 [hep-th]].
  

\bibitem{Chachamis:2011rw}
  G.~Chachamis, M.~Deak, A.~Sabio Vera and P.~Stephens,
  Nucl.\ Phys.\ B {\bf 849} (2011) 28
  [arXiv:1102.1890 [hep-ph]].

\bibitem{Chachamis:2015zzp} 
  G.~Chachamis and A.~Sabio Vera,
  arXiv:1511.03548 [hep-ph].
 
 
\bibitem{Bury:2015dla} 
  M.~Bury and A.~van Hameren,
  Comput.\ Phys.\ Commun.\  {\bf 196}, 592 (2015)
  doi:10.1016/j.cpc.2015.06.023
  [arXiv:1503.08612 [hep-ph]].
 
\bibitem{Nefedov:2013ywa} 
  M.~A.~Nefedov, V.~A.~Saleev and A.~V.~Shipilova,
  Phys.\ Rev.\ D {\bf 87}, no. 9, 094030 (2013)
  doi:10.1103/PhysRevD.87.094030
  [arXiv:1304.3549 [hep-ph]].
  
\bibitem{vanHameren:2012uj} 
  A.~van Hameren, P.~Kotko and K.~Kutak,
  JHEP {\bf 1212}, 029 (2012)
  doi:10.1007/JHEP12(2012)029
  [arXiv:1207.3332 [hep-ph]].
 
\bibitem{Maciula:2015}
 R.~Maciula and A.~Szczurek,
  Phys.\ Lett.\ B {\bf 749} (2015) 57
  [arXiv:1503.08022 [hep-ph]].
 
\bibitem{Maciula:2014pla} 
  R.~Maciula and A.~Szczurek,
  Phys.\ Rev.\ D {\bf 90}, no. 1, 014022 (2014)
  doi:10.1103/PhysRevD.90.014022
  [arXiv:1403.2595 [hep-ph]].




\end{thebibliography}
\end{document}